\documentclass{emulateapj-rtx4}
\usepackage{graphicx}
\usepackage{epsf}
\usepackage{epsfig}
\usepackage{amssymb,amsmath}
\usepackage[usenames]{color}
\usepackage{amssymb}
\usepackage{times}
\usepackage{natbib}
\newcommand{\cf}{\textit{cf.}~}
\newcommand{\ie}{\textit{i.e.}~}
\newcommand{\eg}{\textit{e.g.}~}

\usepackage[T1]{fontenc}
\fontfamily{cmr}

\def\wh{\widehat}

\begin{document}

\title{Black hole-neutron star mergers and short GRBs: a relativistic
  toy model to estimate the mass of the torus}

\author{Francesco Pannarale\altaffilmark{1},
  Aaryn Tonita\altaffilmark{1},
  Luciano Rezzolla\altaffilmark{1,2}}

\altaffiltext{1}{Max-Planck-Institut f{\"u}r Gravitationsphysik,
  Albert Einstein Institut, Potsdam, Germany}
\altaffiltext{2}{Department of Physics and Astronomy, Louisiana State
  University, Baton Rouge, LA, USA}


\begin{abstract}
  The merger of a binary system composed of a black hole and a neutron
  star may leave behind a torus of hot, dense matter orbiting around
  the black hole. While numerical-relativity simulations are necessary
  to simulate this process accurately, they are also computationally
  expensive and unable at present to cover the large space of possible
  parameters, which include the relative mass ratio, the stellar
  compactness, and the black hole spin. To mitigate this and provide a
  first reasonable coverage of the space of parameters, we have
  developed a method for estimating the mass of the remnant torus from
  black hole-neutron star mergers. The toy model makes use of an
  improved relativistic affine model to describe the tidal
  deformations of an extended tri-axial ellipsoid orbiting around a
  Kerr black hole and measures the mass of the remnant torus by
  considering which of the fluid particles composing the star are on
  bound orbits at the time of the tidal disruption. We tune the toy
  model by using the results of fully general-relativistic simulations
  obtaining relative precisions of a few percent and use it to
  extensively investigate the space of parameters. In this way we find
  that the torus mass is largest for systems with highly spinning
  black holes, small stellar compactnesses, and large mass ratios. As
  an example, tori as massive as $M_{b,\text{tor}} \simeq
  1.33\,M_{\odot}$ can be produced for a very extended star with
  compactness $C\simeq 0.1$ inspiralling around a black hole with
  dimensionless spin $a=0.85$ and mass ratio $q\simeq 0.3$.  However,
  for a more astrophysically reasonable mass ratio $q \simeq 0.14$ and
  a canonical value of the stellar compactness $C\simeq 0.145$, the
  toy model sets a considerably smaller upper limit of
  $M_{b,\text{tor}} \lesssim 0.34\,M_{\odot}$.
\end{abstract}

\maketitle

\section{Introduction}\label{sec:intro}

The most widely accepted scenario to explain the phenomenology
associated with short-hard gamma-ray bursts (SGRBs) involves the
merger of either black hole (BH) neutron star (NS) systems or of
binary NS systems~\citep{Nakar:2007yr}. In either case, the remnant
consists of a BH with negligible baryon contamination along its polar
symmetry axis and of a hot, massive accretion torus surrounding it,
which releases energy as it accretes onto the black hole, typically in
the form of a relativistic jet. With these fundamental ingredients of
the standard SGRB model, an intense neutrino flux is emitted as the
torus accretes onto the BH, triggering a high-entropy gas outflow off
the surface of the accretion torus, \ie ``neutrino wind''. At the same
time, energy deposition by $\nu\bar{\nu}$ annihilation in the
baryon-free funnel around the BH rotation axis powers relativistically
expanding $e^\pm$ jets, which can give rise to the observed GRB
emission. Other burst mechanisms have, of course, been proposed; since
these principally involve magnetically launched jets and since we do
not address magnetic fields in this paper, we have briefly summarised
only the burst mechanism powered by neutrino annihilation; the
interested reader may refer to \citet{Lee:2007js} for a thorough
review.

The simulation of these events \textit{``ab initio''} requires an
adequate description of general relativity, relativistic
(magneto)hydrodynamics, and a proper microphysical equation of
state. Typically, the only way to model these systems accurately is to
resort to numerical-relativity simulations, solving consistently both
the Einstein equations and those of relativistic hydrodynamics or
magnetohydrodynamics. These simulations have made considerable
progress in the last few years (see, for
instance~\citet{Oechslin07b,Anderson2007, Baiotti08, Yamamoto2008,
  Anderson2008, Liu:2008xy, Giacomazzo:2009mp, Rezzolla:2010,
  Bauswein:2010dn} for recent studies of NS-NS binaries
or~\citet{Kluzniak99c,Rosswog05, Loeffler06a, Etienne2007b,
  ShibataTaniguchi2008, Duez09,
  Etienne:2008re,Duez:2008rb,Shibata:2009cn,Chawla:2010sw} for
corresponding work on BH-NS binaries). Despite the fact that this type
of simulations is now possible, they remain nevertheless both
challenging and computationally intensive. Numerical simulations of
NS-NS mergers have now reached a rather high level of accuracy
(see~\citet{Baiotti:2009gk} and the discussion in the Appendix
of~\citet{Rezzolla:2010}), and different codes have been shown to
yield results that agree to $10\%$ (at worse) when using the same
initial data~\citep{Baiotti:2010ka}. However, the situation is much
more uncertain in the case of BH-NS binaries, for which no direct
comparison among different codes has been made yet and the results of
the simulations from different codes are sometimes not in
agreement. As an example, the merger of the same binary with mass
ratio $1/3$ yields a torus with baryon mass which is $\sim 4\%$ of the
NS in \citet{Etienne:2008re} and $\lesssim 0.001\%$
in~\citet{Shibata:2009cn}. As a result, no reliable knowledge is
available at the moment on how the mass of the torus depends on the
most important parameters of the system: the mass ratio, the stellar
compactness, and the BH spin.

These problems, along with the need of a better understanding of the
tidal disruption process, have pushed the parallel development of
pseudo-Newtonian BH-NS calculations -- \eg \citet{Ruffert2010} use the
Paczy\'nski-Wiita phenomenological potential to mimic the innermost
stable circular orbit (ISCO) of the BH in a Newtonian setting -- and
of semi-analytical approaches to the problem. Regarding the latter,
\citet{Shibata96}, for instance, described the necessary conditions
for the production of an accretion torus of appreciable size by
requiring that the NS disruption occurs at a tidal radius $r_{\rm
  tide}$ that is larger than the ISCO of the BH
$r_{\text{ISCO}}$. Unfortunately, \citet{Shibata96} did not predict
the mass of the resulting accretion torus except to assume that it
vanishes when the radius of tidal disruption is less than that of the
ISCO. A parallel systematic study has been pursued recently to exploit
the relativistic \textit{``affine-model''} and describe the properties
of the tidally deformed NS~\citep{Ferrari09,Ferrari:2009bw}. Also in
this case, however, the study has concentrated on the evolution of
stationary configuration and has not made any prediction on the final
outcome of the merger.

In this paper we attempt to bridge the gap between intensive numerical
simulations and semi-analytical studies by establishing a way to
estimate the mass of the resulting torus. We do this by taking the
concept of the tidal disruption to its logical extreme. In other
words, we model the NS in the binary as a relativistic tri-axial
ellipsoid which is tidally distorted as it orbits in the tidal field
of a rotating BH. When the tidal-disruption radius is reached,
however, we assume the star to be composed of a system of
non-interacting ``fluid particles'' which move on the corresponding
geodesics. We therefore compute the mass of the torus as the integral
of the masses of the particles which do not fall into the BH. This
clearly represents only a \textit{``toy model''} for the complex
dynamics of the merger process, but we show that, with a suitable
tuning, it allows us to reproduce with good precision the large
majority of the results obtained so far from more accurate but also
considerably more expensive numerical relativity calculations. Most
importantly, however, it provides a simple tool to better understand
the complex dynamics of the tidal disruption and to cover at once the
full space of parameters.

The structure of the paper is the following one. In
Sect.~\ref{sec:method} we describe the particular tidal model we use
and then how we estimate the mass of the accretion torus. In
Sect.~\ref{sec:tuning} we show that by tuning the free parameter in
our model we can reproduce results obtained within fully
general-relativistic simulations, thus proving that the tool we build
is solid. In Sect.~\ref{sec:results} we present the results of our
estimates, leaving an intuitive interpretation of the results and the
conclusive overview to Sects.~\ref{sec:intepretation} and
~\ref{sec:concl}, respectively.

\section{Method}\label{sec:method}

To model the behaviour of the NS during the final stages of the
inspiral of the mixed binary and before it merges with the BH, we use
an improved version of the affine model which is presented in detail
in~\citet{Ferrari09}. An important difference with respect to that
work is that we do not consider the prescriptions of the
quasi-equilibrium approximation and, rather, follow the dynamics of
the NS until it is disrupted by the BH tidal field. Furthermore, as
mentioned in the Introduction, in addition to treating the NS as a
tri-axial ellipsoid, we also decompose it into a large number of
representative ``fluid particles'', whose kinematic properties will be
used to study the motion of the NS matter after the tidal
disruption. Our toy model is therefore composed of three logical
parts: \textit{(i)} the evolution of the NS deformation as it
inspirals towards the BH; \textit{(ii)} the modelling of the tidal
disruption; \textit{(iii)} the calculation of the mass building-up the
torus. Each of these parts will be discussed separately in the
remainder of this Section.

\subsection{Neutron star deformation - The affine model}\label{subsec:aff-model}

The idea of modelling stars as ellipsoids has a long history and a
thorough analysis of incompressible ellipsoidal figures of equilibrium
was performed by Chandrasekhar in 1969~\citep{Chandrasekhar69c}. When
modelling the NS deformation, we are addressing what is known as the
\emph{compressible Roche-Riemann problem}, in which one studies the
behaviour of a compressible ellipsoid with uniform vorticity parallel
to its rotation axis, orbiting a point mass or a rigid sphere. More
specifically, we will be using an improved version of the affine
model, which was developed in the '80s by Carter, Luminet and Marck to
describe the encounters between a BH and a Newtonian
star~\citep{Carter82,Carter83,Carter85,Luminet85,Carter86} and then
applied to BH-NS binaries at the very end of the
'90s~\citep{Wiggins00}. More recently, the Newtonian treatment of the
star was upgraded to achieve a better description of the NS in mixed
compact binaries~\citep{Ferrari09,Ferrari:2009bw}. The essential
features and assumptions of the improved affine approach used here can
be summarized as follows:

\begin{itemize}

\item the equilibrium structure of the NS is determined by the TOV
  equations, while its dynamical behaviour is governed by Newtonian
  hydrodynamics improved by the use of an effective relativistic
  scalar potential~\citep{Rampp02};

\item the NS centre of mass moves in the tidal field of a Kerr BH
  along a simple inspiralling equatorial orbit [\cf
  eq.~(\ref{eq:orbit})] and each point of the orbit is associated with
  a BH timelike circular geodesic having the same radius;

\item throughout the inspiral, the NS remains a Riemann S-type
  ellipsoid, \ie its spin and vorticity are parallel and their ratio
  is constant (see~\citet{Chandrasekhar69c});

\item tidal effects on the orbital motion and the perturbation that
  the star induces on the BH are neglected.

\end{itemize}

\begin{figure}[t]
  \begin{center}
    \includegraphics[width=8.0cm,angle=0]{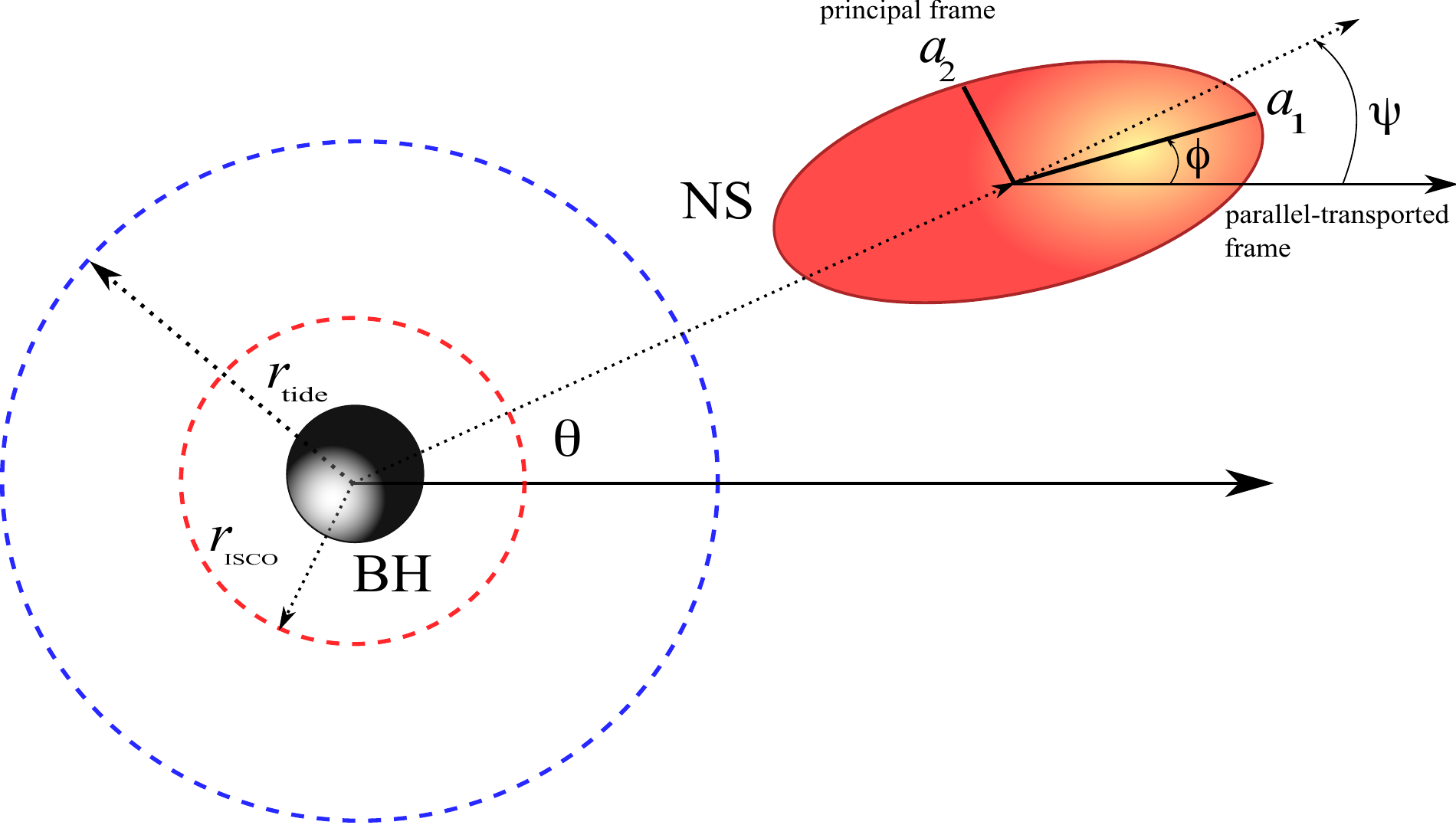}
    \vskip 1.0cm
    \caption{\label{cartoon}Cartoon of the toy model. Indicated are
      the tidal radius $r_{\rm tide}$, the ISCO $r_{\text{ISCO}}$, two
      of the principal axes $a_1, a_2$, the principal frame and the
      parallel-transported frame. Note that for simplicity we set
      $\phi=\Psi$.}
  \end{center}
\end{figure}

For completeness, we next review the mathematical formulation of the
affine model used here by writing the equations governing the NS
deformations in the principal frame, \ie the frame associated with the
principal axes of the stellar ellipsoid. In this frame, the fluid
variables of the affine model are five: the three principal axes of
the stellar ellipsoid $a_1,a_2$, and $a_3$, the angular frequency of
the internal fluid motion $\Lambda$ and the star spin $\Omega$
measured in the parallel-transported frame associated with the centre
of mass of the star~\citep{Marck83}. The axis $a_3$ is perpendicular
to the orbital plane, while $a_1$ and $a_2$ are perpendicular to $a_3$
(see Fig.~\ref{cartoon}). In the Newtonian limit ($M_\text{BH}\ll r$,
where $M_\text{BH}$ is the BH mass and $r$ is the Boyer-Lindquist
radial coordinate), $a_1$ and $a_2$ belong to the orbital plane. The
dynamics of four of the five fluid variables is then governed by the
following set of equations
\begin{align}
\ddot a_1 &= a_1(\Lambda^2+\Omega^2) - 2a_2\Lambda\Omega
+ \frac{1}{2}\frac{\hat{V}}{\wh{\mathcal{M}}}R_{\text{NS}}^3a_1\tilde{A}_1
+ \frac{R_{\text{NS}}^2}{\wh{\mathcal{M}}}\frac{\Pi}{a_1}\nonumber\\
\label{dda1}
 &- c_{11}a_1\,,\\
\ddot a_2 &= a_2(\Lambda^2+\Omega^2) - 2a_1\Lambda\Omega
+ \frac{1}{2}\frac{\hat{V}}{\wh{\mathcal{M}}}R_{\text{NS}}^3a_2\tilde{A}_2
+ \frac{R_{\text{NS}}^2}{\wh{\mathcal{M}}}\frac{\Pi}{a_2}\nonumber\\
\label{dda2}
 &- c_{22}a_2\,,\\
\label{dda3}
\ddot a_3 &= \frac{1}{2}\frac{\hat{V}}{\wh{\mathcal{M}}}R_{\text{NS}}^3a_3
\tilde{A}_3 + \frac{R_{\text{NS}}^2}{\wh{\mathcal{M}}}\frac{\Pi}{a_3}
- c_{33}a_3\,,\\
\label{dJs}
\dot{J}_s &= \frac{\wh{\mathcal{M}}}{R_{\text{NS}}^2}c_{12}(a_2^2-a_1^2)\,,
\end{align}
where the dots denote derivatives with respect to the proper time at
the stellar centre $\tau$, the $c_{ij}$'s denote the components of the
BH tidal tensor in the principal frame, $R_{\text{NS}}$ is the NS
radius, and the index symbols $\tilde{A}_i$ are defined
as\footnote{The $\tilde{A}_i$'s are related to the dimensionless index
  coefficients defined in~\citet{Chandrasekhar69c} by the simple
  dimensional rescaling $A_i=R_\text{NS}^5\tilde{A}_i$. In essence,
  they express the derivative of the self-gravity of the deformed star
  with respect to its $i$-th axis.}
\begin{eqnarray}
\label{def:tildeA}
\tilde{A}_i \equiv \int_0^\infty\frac{d\sigma}{(a_i^2+\sigma)
  \sqrt{(a_1^2+\sigma) (a_2^2+\sigma)(a_3^2+\sigma)}}\,.
\end{eqnarray}
The effective relativistic self-gravity potential for the isolated NS
in spherical equilibrium equilibrium $\wh{V}$ and the scalar
quadrupole moment for the isolated NS in spherical equilibrium
equilibrium $\wh{\mathcal{M}}$ are given by\footnote{Hats ( $\hat{}$ )
  denote quantities calculated for an isolated nonrotating NS at
  equilibrium.}
\begin{align}
\label{def:hatV}
\widehat{V}&\equiv-4\pi \int_0^{R_{\text{NS}}}
\frac{[\hat{\epsilon}(\hat{r})+\hat{p}(\hat{r})][m_{\text{TOV}}(\hat{r})
  +4\pi\hat{r}^3\hat{p}(\hat{r})]}{
  \hat{\rho}(\hat{r})\hat{r}[\hat{r}-2m_{\text{TOV}}(\hat{r})]}
\hat{r}^3\hat{\rho} d\hat{r}\,,\\
\label{def:momquad}
\widehat{\mathcal{M}} &\equiv
\frac{4\pi}{3}\int_0^{R_{\text{NS}}}\hat{r}^{4}\hat{\rho}d\hat{r}\,,
\end{align}
with $dm_{\text{TOV}}/dr = 4\pi r^2\epsilon(r)$. The pressure integral $\Pi$
is calculated as
\begin{eqnarray}
\label{def:Pi}
\Pi \equiv \int p(\rho) d^3x = 
4\pi \frac{a_1a_2a_3}{R_{\text{NS}}^3}\int_0^{R_{\text{NS}}}p\left(
  \frac{\hat{\rho}}{a_1a_2a_3}\right)\hat{r}^2d\hat{r}\,,\nonumber \\
\end{eqnarray}
while 
\begin{align}
\label{def:J_s}
J_s &\equiv \frac{\wh{\mathcal{M}}}{R_{\text{NS}}^2}
[(a_1^2+a_2^2)\Omega-2a_1a_2\Lambda]\,,
\end{align}
is the spin angular momentum of the star. The fifth fluid variable may
be expressed in terms of
\begin{align}
\label{def:Circ}
\mathcal{C} &\equiv \frac{\wh{\mathcal{M}}}{R_{\text{NS}}^2}
[(a_1^2+a_2^2) \Lambda-2a_1a_2\Omega]\,,
\end{align}
which is proportional to the circulation in the locally nonrotating
inertial frame. We note that because we work in absence of viscosity,
the circulation of the fluid is conserved, \ie
$\dot{\mathcal{C}}=0$. For simplicity, but also because this is the
assumption made by all numerical-relativity simulations to date, we
set $\mathcal{C}=0$ initially (the NS fluid is thus irrotational) so
that \eqref{def:Circ} reduces to
\begin{eqnarray}
\label{eq:Lambda}
\Lambda = \frac{2a_1a_2\Omega}{a_1^2+a_2^2}\,.
\end{eqnarray}

The components of the tidal tensor for a Kerr spacetime, expressed in
the NS principal frame, are
\begin{align}
\label{eq:c11}
c_{11} &= \frac{M_\text{BH}}{r^3}\left[1-3\frac{r^2+K}{r^2}\cos^2(\Psi-\phi)
\right]\,,\\
c_{22} &= \frac{M_\text{BH}}{r^3}\left[1-3\frac{r^2+K}{r^2}\sin^2(\Psi-\phi)
\right]\,,\\
\label{eq:c12}
c_{12} &= c_{21} = \frac{M_\text{BH}}{r^3}\left[ -\frac{3}{2}\frac{r^2+K}{r^2}
  \sin 2(\Psi-\phi)\right]\,,\\
\label{eq:c33}
c_{33} &= \frac{M_\text{BH}}{r^3}\left( 1+3\frac{K}{r^2}\right)\,,
\end{align}
where the angle $\phi$ (which is related to $\Omega$ by $\dot\phi
\equiv \Omega$) is the angle that brings the parallel-transported
frame into the principal frame by a rotation around the $a_3$
axis~\citep{Marck83}. Similarly, $\Psi$ is an angle that governs the
rotation of the parallel-transported tetrad frame in order to preserve
the parallel transport of its basis vectors. Stated differently, the
difference between $\Psi$ and $\phi$ represents the lag angle between
the principal frame and the parallel transported one, \ie
$\phi_{\text{lag}}\equiv \Psi - \phi$, and thus measures how much the
star is ``lagging behind'' in its orbit around the BH. If
$\phi_{\text{lag}}=0$ and thus $\Psi = \phi$, then the largest
semi-major axis of the star is always pointing towards the BH. A
schematic diagram of the BH-NS binary and of the relevant quantities
discussed so far is shown in Fig.~\ref{cartoon}.

The constant $K$ appearing in the tidal-tensor
components~\eqref{eq:c11}--\eqref{eq:c33} is a combination of the
energy $E$ and the $z$-orbital angular momentum per unit mass of the
star $L_z$
\begin{eqnarray}
K \equiv ({\tilde a}E-L_z)^2\,,
\end{eqnarray}
where ${\tilde a}\equiv J/M$ is the spin of the BH and, for circular
geodesics,
\begin{align}
\label{circularE} E &\equiv \frac{r^2-2rM_\text{BH}+{\tilde a}\sqrt{rM_\text{BH}}}
{r\sqrt{r^2-3rM_\text{BH}+2{\tilde a}\sqrt{rM_\text{BH}}}}\,,\\
\label{circularL} L_z &\equiv \frac{\sqrt{rM_\text{BH}}(r^2-2{\tilde a}
\sqrt{rM_\text{BH}}+{\tilde a}^2)}{r\sqrt{r^2-3rM_\text{BH}+2{\tilde a}\sqrt{rM_\text{BH}}}}\,.
\end{align}
Note that because the tidal-tensor components $c_{11}$ and $c_{22}$
have a different sign, the forces acting on the corresponding
semi-major axes $a_1$ and $a_2$ also have opposite signs and thus lead
to a stretching of $a_1$ and a to compression of $a_2$.

For simplicity, and to obtain a better agreement with the results of
the numerical-relativity simulations, we will consider hereafter
$\phi_{\text{lag}}=0$ and thus $\phi=\Psi$. Furthermore, since we will
consider a sequence of circular equatorial geodesics whose radius
reduces due to the emission of gravitational radiation, the evolution
of the angle $\Psi$ for each circular orbit is given
by~\citep{Marck83}
\begin{eqnarray}
\label{eq:Psi}
\dot{\Psi} = \sqrt{\frac{M_\text{BH}}{r^3}}\,,
\end{eqnarray}
and thus also
\begin{eqnarray}
\label{eq:Omega}
\dot{\phi}=\Omega=\sqrt{\frac{M_\text{BH}}{r^3}}\,.
\end{eqnarray}

In order to evolve the equations of the affine
model~\eqref{dda1}--\eqref{dda3} we must select an equation of state
(EOS) for the NS matter and specify the initial conditions and the
evolution of the orbit. As far as the first is concerned, the model is
sufficiently general that any EOS could be used and indeed several
different ones were used in~\citet{Ferrari:2009bw}. However, because
here we want to compare with the results of numerical-relativity
simulations and these have been performed mostly with a $\Gamma$ law
equation of state $p=(\Gamma-1)\rho\epsilon$ which for the adiabatic
process considered in this paper is equivalent to a polytropic EOS
$p=K\rho^{\Gamma}$ with $\Gamma=2$ or $2.75$, we will consider here
just polytropes with these polytropic exponents and present the result
of more realistic EOSs in a subsequent work. As for the initial
conditions, we consider an initial separation for the binary,
$r_0\equiv r(t=0)$ and set $\phi_0\equiv \phi(t=0)=0$, while $\Omega$
and $\Lambda$ are automatically given by~\eqref{eq:Omega}
and~\eqref{eq:Lambda}, respectively. For the NS axes, instead, we set
the time derivatives on the left hand sides
of~\eqref{dda1}--\eqref{dda3} to zero and solve the system for
$a_1,a_2,a_3$ with a Newton-Raphson scheme. Of course, it is necessary
to ensure that $r_0$ is large enough, \ie that initially $a_1\simeq
a_2\simeq a_3\simeq R_{\text{NS}}$ and $J_s\simeq 0$ (as well as
$\mathcal{C}=0$), and that the final results are unchanged if one
chooses a larger $r_0$. In other words, at $t=0$ the NS must almost be
at spherical equilibrium and the calculations must therefore be
independent of the specific choice made for $r_0$.

For the time evolution of the orbital separation $r$, we consider a
very simple circular equatorial adiabatic inspiral~\citep{Misner73},
which accounts therefore for the radiative losses of two point masses
at a 2.5 post-Newtonian (PN) approximation
\begin{eqnarray}
\label{eq:orbit}
r(t) = r_0\left( 1-\frac{t}{t_c}\right)^{1/4}\,,
\end{eqnarray}
where
\begin{eqnarray}
t_c = \left(\frac{5}{256}\right)
\frac{r_0^4}{M_{\text{NS}}M_\text{BH}(M_{\text{NS}}+M_\text{BH})}\,,
\end{eqnarray}
is the inspiral time and $M_{\text{NS}}$ is the gravitational mass of
the NS. When generating the orbit, we evolve the orbit angle $\theta$
according to the Kerr spacetime equation
\begin{align}
\label{eq:dthetadt}
\frac{d\theta}{dt} &= \frac{1}{\tilde{a}+\sqrt{r^3/M_\text{BH}}}
\end{align}
and we make the approximation of setting
\begin{align}
\label{eq:dtdtau}
\frac{d\tau}{dt} &= 1\,.
\end{align}
It is important to remark that our goal is that of computing the mass
of the torus produced by the tidal disruption and not that of
providing an accurate description of the binary inspiral. In this
sense, using a lower-order PN description of the orbit is very
reasonable as the dynamics we are most interested in take place when
the presently available PN models are no longer accurate.

Once the orbit is determined, we may integrate the affine-model
equations, terminating the evolution when the ratio of the semi-major
axes reaches a critical value $(a_2/a_1)_{\text{crit}}$. This quantity
cannot be determined a priori and is effectively a free parameter in
our toy model. However, it may be tuned by comparing the results of
the toy model with those of the numerical simulations and the way we
do this will be discussed in the next Section. We thus define the
tidal disruption radius $r_{\text{tide}}$ as the orbital separation at
which $(a_2/a_1)=(a_2/a_1)_{\text{crit}}$.

A final quantity which is relevant to introduce and that may be useful
to interpret the results of the toy model is the ISCO, which, for a
generic Kerr BH is given by~\citep{Bardeen72}
\begin{align}
r_{\text{ISCO}} &= M_\text{BH}\{3+Z_2\mp [(3-Z_1)(3+Z_1+2Z_2)]^{1/2}\}\,,\nonumber\\
Z_1 &= 1+(1-{\tilde a}^2/M_\text{BH}^2)^{1/3}\,,\nonumber\\
&\times[(1+{\tilde a}/M_\text{BH})^{1/3}+
(1-{\tilde a}/M_\text{BH})^{1/3}]\,,\nonumber\\
Z_2 &= (3{\tilde a}^2/M_\text{BH}^2+Z_1^2)^{1/2}\,,
\label{def:risco}
\end{align}
where the upper/lower sign holds for co-rotating/counter-rotating
orbits. In general, the ISCO is inside the tidal radius, \ie
$r_{\text{ISCO}} < r_{\text{tide}}$, but there are situations in which
the opposite is true and this is the case, for instance, when
considering binary systems with very small mass ratios or stars with
very large compactness. In these cases too, we follow the evolution of
the axis ratio and ``disrupt'' the NS inside the ISCO as soon as the
critical value is reached.

\subsection{Neutron star disruption}\label{subsec:NSdisruption}

As mentioned above, when the affine-model evolution of the mixed
binary leads to the tidal disruption of the NS, we fragment the NS
into fiducial fluid elements that would be representative of the
motion of the NS matter.  The first step in our strategy consists
therefore in switching from the five fluid variables of the
affine-model formulation to a description of the (disrupted) NS fluid
as a set of test particles, each one of which possesses a mass and a
$4$-velocity. In practice, at disruption we build a fine grid adapted
to the ellipsoidal shape of the star and divide the star into a
collection of fluid elements. In the principal frame, the centre of
each fluid element is identified by a $3$-vector $\vec{x}$ and we
calculate the mass of the corresponding fluid cell by multiplying the
mass density at its centre by the volume of the fluid element.
Moreover, we may associate to the centre of mass of each cell a
$3$-velocity, which, in the principal-axes frame, is given
by~\citep{Chandrasekhar69c}
\begin{eqnarray}
\vec{u} =\vec{u}_s + \vec{u}_e\,,
\end{eqnarray}
where
\begin{eqnarray}
\vec{u}_s \equiv \frac{a_1}{a_2}\Lambda x_2\vec{e}_1
-\frac{a_2}{a_1}\Lambda x_1\vec{e}_2\,,
\end{eqnarray}
is the spin velocity (\ie the speed of the fluid due to its rotation),
and
\begin{eqnarray}
\vec{u}_e \equiv \frac{\dot{a}_1}{a_1}x_1\vec{e}_1
+\frac{\dot{a}_2}{a_2}x_2\vec{e}_2+
\frac{\dot{a}_3}{a_3}x_3\vec{e}_3\,,
\end{eqnarray}
is the ellipsoid expansion/contraction velocity, $\vec{e}_i$ being the
unit vectors along the ellipsoid principal axis $a_i$. The coordinate
$x_i$ along the $i$-th axis runs from $-a_i$ to $a_i$. With a rotation
of an angle $\phi$ around $a_3$, we switch from position $3$-vectors
in the principal frame to position $3$-vectors in the
parallel-transported tetrad associated with the NS centre of mass,
where the $3$-velocity $\vec{u}$ becomes $\vec{u} + \vec{\Omega}
\times \vec{x}$. In this parallel-transported tetrad, we determine the
time component of each $4$-velocity vector by exploiting the
normalization condition $u^{(\alpha)}u_{(\alpha)}=-1$ and by recalling
that the time component of the position vectors is simply
$x^{(0)}=0$. Finally, we express all the $4$-position and the
$4$-velocity vectors in Boyer-Lindquist coordinates by applying the
transformation laws derived in~\citet{Marck83} and summarised in
Appendix~\ref{app:parallel-tetrad}.

The procedure described above provides a complete description, in
Boyer-Lindquist coordinates, of the kinematic properties of fluid
parcels as point particles freely-falling in a Kerr spacetime; this is
what is needed to then estimate the torus mass.

\subsection{Torus mass estimation}\label{subsec:TorusMass}

When looking carefully in numerical-relativity simulations at the
dynamics of the NS after it is disrupted, it is quite striking to note
how much the different parts of the star seem to behave like
independent freely falling particles: the gravity of the black hole
alone does seem to represent the dominant force at this stage of the
evolution. In view of this observation, when the NS is tidally
disrupted and split into fiducial fluid elements of which we know the
mass and the $4$-velocity, we assume that the pressure gradients
across neighbouring elements and the self-gravity of the system play
little role, and hence that the fluid elements behave as independent
collisionless fluid particles. As such, after the disruption the NS is
approximated as an ensemble of about $3.1\times 10^{4}$ fluid
particles which have a complex distribution of energy and angular
momenta, but are in free-fall towards the BH.

Using the $4$-velocity, of each particle we compute the
corresponding conserved quantities by inverting the relations
\begin{align}
\frac{dt}{d\tau} &= \frac{-{\tilde a}({\tilde a}e\sin^2\theta - \ell_z) + 
(r^2 + {\tilde a}^2)P/\Delta}{r^2 + {\tilde a}^2\cos^2\theta}\,,\\
\label{radialEquation}
\left(\frac{dr}{d\tau}\right)^2 &= \frac{P^2 - 
\Delta[r^2 + (\ell_z - {\tilde a}e)^2 + {\cal Q}]}{(r^2 + 
{\tilde a}^2\cos^2\theta)^2}\,,\\
\left(\frac{d\theta}{d\tau}\right)^2 &= 
\frac{{\cal Q} - \cos^2\theta[{\tilde a}^2(1-e^2) + \ell_z^2/\sin^2\theta]}{(r^2 + 
{\tilde a}^2\cos^2\theta)^2}\,,\\
\frac{d\phi}{d\tau} &= \frac{-({\tilde a}e - \ell_z/\sin^2\theta) + 
{\tilde a}P/\Delta}{r^2 + {\tilde a}^2\cos^2\theta}\,,
\end{align}
where 
\begin{align}
P &\equiv e(r^2 + {\tilde a}^2) - \ell_z{\tilde a}\,,\\
\Delta &\equiv r^2 - 2rM_\text{BH} + {\tilde a}^2\,,
\end{align}
and $e$, $\ell_z$, and ${\cal Q}$ represent the energy, angular
momentum and Carter's constant of motion, respectively, all normalized
to the mass of the particle. Note that it is necessary to use these
general equations instead of equations~\eqref{circularE}
and~\eqref{circularL} as the majority of the particles no longer
follows circular equatorial geodesics.

As mentioned previously, we identify the mass of the remnant torus
with the sum of the masses of the bound particles and we make use
of~\eqref{radialEquation} to determine whether a given particle is
bound or not. Noting that a turning point occurs when $(dr/d\tau)^2$
passes through zero and since the only influence of $\theta$ is to
decrease the overall magnitude of $(dr/d\tau)^2$ but not to change its
sign, we only consider, without loss of generality, the case
$\theta=\pi/2$. We then use root-finding techniques for each particle
and consider bound those particles for which $(dr/d\tau)^2<0$ at a
radial position $r_\text{TP}$ outside the event horizon
$r_{\text{EH}}$, such that $r_\text{EH} < r_\text{TP} < r_\text{tide}$
(note that $(dr/d\tau)^2$ is always greater than $0$ at
$r=r_\text{tide}$) and simultaneously satisfy $e<1$. This final
condition merely states that the gravitational binding energy has
compensated the kinetic energy such that the total energy of the
particle is less than the rest mass of the particle at infinity.

Once the NS is tidally disrupted, the calculation of the torus mass,
which is initially set to be $M_{\rm b,tor}=M_{\rm b, NS}$, is done as
follows.
\begin{enumerate} 
\item For each fluid particle we verify whether it is bound or
  not. In this latter case we assume the particle will accrete
  onto the BH\footnote{We note that all numerical simulations suggest
    that the amount of matter leaving the central gravitational
    potential, \ie that are unbounded but do not fall onto the BH, is
    extremely small and can thus be neglected here
    (see~\citet{Rezzolla:2010}).}.
\item The composite mass of the accreted particles is added to the mass of
  the BH and the mass of the torus is decreased by the corresponding
  amount.
\item We reconsider the remaining particles and verify if they are
  still bound or if they would now accrete onto the new and more
  massive BH.
\end{enumerate}
This procedure is repeated until there are no more particles that would
accrete onto the BH or, equivalently, until the relative change
in the mass of the torus is less than one part in one million.

In addition to a change in the mass of the BH we have also
experimented with changing the spin of the black hole as a result of
the angular momentum accreted with the particles. However, the results
in this case are much less robust (the mass of the torus is not a
monotonic function of the parameters) and this is probably due to the
more complex dependence of the geodesic motion on the spin of the BH,
which conflicts with the approximations made here. As a result, we
keep the BH spin to be the same as the initial one and it is
reassuring that this does not spoil the very good agreement with the
numerical simulations.

\begin{table*}
  \caption{Comparison between the remnant torus mass predictions of
    fully general-relativistic simulations and of our model with the
    critical value of $a_2/a_1$ tuned to $0.44$. In the four sections
    of the table, we examine the results recently provided in (from
    top to bottom)~\citet{Tonita:2010}, \citet{Duez09}, \citet{Etienne:2008re},
    and~\citet{Shibata:2009cn}. The first four columns of the table
    are the parameters of each BH-NS binary, \ie the adiabatic index
    $\Gamma$ of the NS EOS, the NS compactness $C$, the mass ratio $q$
    and the dimensionless BH spin $a$. The following three columns
    provide the remnant torus masses $M_\text{tor}$ obtained from our
    model (labelled \textit{``toy model''}), those obtained from fully
    general-relativistic calculations (labelled
    \textit{``simulations''}), both given in units of the NS baryonic
    mass $M_{b,\text{NS}}$, and the relative error.\label{tab:tune}}
\begin{center}
\begin{tabular}{|ccccc|rrr|}
\hline
\hline
Ref.& EOS         & $C$ & $q$ & $a$ &  $M_{b, \text{tor}}/M_{b, \text{NS}}$ & $M_{b, \text{tor}}/M_{b, \text{NS}}$ & error \\
    & ($\Gamma$)  &             &    &     & (\textit{toy model})          & (\textit{simulations})          & ($\%$)\\
\hline
\citet{Tonita:2010}    & $2.00$ & $0.100$ & $1/5$   & $~~~0.00$ & $0.17$  & $0.17$  & $0  $ \\
\citet{Tonita:2010}    & $2.00$ & $0.125$ & $1/5$   & $~~~0.00$ & $0.06$  & $0.06$  & $0  $ \\
\citet{Tonita:2010}    & $2.00$ & $0.145$ & $1/5$   & $~~~0.00$ & $<0.01$ & $<0.01$ & $0  $ \\
\citet{Tonita:2010}    & $2.00$ & $0.150$ & $1/5$   & $~~~0.00$ & $<0.01$ & $<0.01$ & $0  $ \\
    \hline                                                                         
\citet{Duez09}         & $2.00$ & $0.144$ & $1/3$   & $~~~0.50$ & $0.08$  & $0.08$  & $0  $ \\
\citet{Duez09}         & $2.75$ & $0.146$ & $1/3$   & $~~~0.50$ & $0.11$  & $0.13$  & $18 $ \\
\citet{Duez09}         & $2.75$ & $0.173$ & $1/3$   & $~~~0.50$ & $0.04$  & $0.02$  & $50 $ \\
    \hline                                                                         
\citet{Etienne:2008re} & $2.00$ & $0.145$ & $1/3$   & $~~~0.00$ & $0.02$  & $0.04$  & $100$ \\
\citet{Etienne:2008re} & $2.00$ & $0.145$ & $1/3$   & $~~~0.75$ & $0.18$  & $0.15$  & $17 $ \\
\citet{Etienne:2008re} & $2.00$ & $0.145$ & $1/3$   & $-0.50$   & $<0.01$ & $<0.01$ & $0  $ \\
\citet{Etienne:2008re} & $2.00$ & $0.145$ & $1/5$   & $~~~0.00$ & $<0.01$ & $<0.01$ & $0  $ \\
\tableline                                                                             
\citet{Shibata:2009cn} & $2.00$ & $0.145$ & $1/3$   & $~~~0.00$ & $0.02$  & $<0.01$ & $100$ \\ 
\citet{Shibata:2009cn} & $2.00$ & $0.160$ & $1/3$   & $~~~0.00$ & $<0.01$ & $<0.01$ & $0  $ \\
\citet{Shibata:2009cn} & $2.00$ & $0.178$ & $1/3$   & $~~~0.00$ & $<0.01$ & $<0.01$ & $0  $ \\
\citet{Shibata:2009cn} & $2.00$ & $0.145$ & $1/4$   & $~~~0.00$ & $0.01$  & $<0.01$ & $100$ \\ 
\citet{Shibata:2009cn} & $2.00$ & $0.145$ & $1/5$   & $~~~0.00$ & $<0.01$ & $<0.01$ & $0  $ \\
\tableline
\tableline
\end{tabular}
\end{center}
\end{table*}

\section{Tuning and validation of the model}\label{sec:tuning}

In the affine-model approach based on a quasi-equilibrium
approximation and discussed in~\citet{Ferrari09}, the disruption
radius is identified by the condition $[\partial (a_2/a_1)/\partial
r]^{-1} =0$,~\ie as the radial separation at which the axis ratio
diverges. Although this singular limit is clearly a shortcoming of the
assumption of quasi-equilibrium, it is not obvious how to specify the
tidal radius in a way which is not arbitrary to some extent. To remove
at least in part this degree of arbitrariness, we have decided to tune
the tidal radius by carefully analyzing the results of recent
numerical-relativity simulations and in particular those carried out
at the AEI~\citep{Tonita:2010}, for which we have more direct control
over the errors. When doing so, we realized that the critical value of
the axis ratio $(a_2/a_1)_{\text{crit}}$ is a robust measure across
our simulations, but also when comparing with the simulations
published in the literature. Hence, we have decided to consider the
critical axis ratio $(a_2/a_1)_{\text{crit}}$ as a free parameter and
to identify its value as the one which best reproduces the numerical
data available.

More specifically, for those initial data for which numerical
simulations have been performed, we tuned, within the toy model, the
choice of the free parameter $(a_2/a_1)_{\text{crit}}$ so as to
minimize the difference, with the corresponding numerical-relativity
result, for the torus mass. As a result of this procedure we we obtain
$(a_2/a_1)_{\text{crit}}=0.44$ which is robust across all of the
simulations and thus define the tidal radius as the orbital separation
at which $(a_2/a_1)$ attains the critical value. It is reasonable to
expect that $(a_2/a_1)_{\text{crit}}$ will depend on the BH spin and
on the mass ratio. Here, however, we assume that such dependence is
weak and thus set it to be be constant. As we discuss below, even with
this crude approximation we can reproduce most of the numerical
results with an error which is below $\sim 15\%$. As an addition note,
we stress that although robust (\ie a single choice fits well all of
the available data), the masses of the tori are rather sensitive to
the choice for the critical axis ratio. In particular, for the same
binary, a change of $\sim 2\%$ in $(a_2/a_1)_{\text{crit}}$ (\ie a
change in the last significant figure) may lead to a change in the
last significant figure of the estimated torus mass, and thus up to a
$\sim 50\%$ change for cases with a very small remnant mass. This
effect disappears if one tunes $(a_2/a_1)_{\text{crit}}$ with an extra
significant digit.

Before going to the details of the comparison with the numerical
simulations it is worth making two remarks. The first one is that
after having identified in the axis ratio a consistent parameter which
we constrain to the second significant figure, we also expect that it
will be further refined as new and more accurate results from
numerical simulations become available. The second one has already
been made in the introduction and stresses the fact that the numerical
data itself does not show a great degree of consistency. While there
are two cases which have been considered by more than one group, most
of the data available refers to configurations which are slightly
different and hence difficult to compare. Even the actual procedure
followed to measure the mass of the tori differs from group to group;
while most decide to measure the mass at a given time after the
formation of the apparent horizon, not all groups use the same
time. It would certainly be more reasonable if the measure was
performed only when the mass accretion rate has reached a very small
and constant value, as done in~\cite{Rezzolla:2010}, rather than
setting a time which may vary from simulation to
simulation. Notwithstanding these difficulties, it is remarkable that
even for the same configurations (\cf the eighth and the twelfth rows
in Table~\ref{tab:tune}), or some which are not very different (\cf
the fifth and ninth rows in Table~\ref{tab:tune}), the numerical
results yield tori whose masses differ considerably. Interestingly,
the predictions of the toy model are equally distant from the
numerical results reported in the eighth and the twelfth rows, thus
suggesting that both simulations may be equally imprecise.

\begin{table*}[t]
  \caption{Gravitational-wave frequency at the onset of tidal
    disruption $f_\text{tide}^\text{GW}$ as computed with our model
    (labelled \textit{``toy model''}) or as quoted
    in~\citet{ShibataTaniguchi2008} (labelled
    \textit{``simulations''}). The latter values were calculated by
    means of a fitting formula determined in \citet{Taniguchi:2008a}
    by using quasi-equilibrium sequences of mixed binaries in circular
    orbits, obtained by solving the Einstein constraint equations in
    the conformal thin-sandwich decomposition and the relativistic
    equations of hydrostationary equilibrium. \label{tab:freq}}
\begin{center}
\begin{tabular}{|cccc|cc|}
\tableline
\tableline
$q$ & $M_{b, \text{NS}}$ [$M_\odot$] & $M_\text{NS}$ [$M_\odot$] & $R_\text{NS}$ [km] & $f_\text{tide}^\text{GW}$ [kHz] & $f_\text{tide}^\text{GW}$ [kHz] \\
 & & & & (\textit{toy model}) & (\textit{simulations}) \\
\hline
$0.327$ & $1.400$ & $1.302$ & $13.2$  & $0.856$ & $0.855$ \\
$0.327$ & $1.400$ & $1.294$ & $12.0$  & $0.997$ & $0.993$ \\
$0.328$ & $1.400$ & $1.310$ & $14.7$  & $0.736$ & $0.738$ \\
$0.392$ & $1.400$ & $1.302$ & $13.2$  & $0.877$ & $0.867$ \\
$0.392$ & $1.400$ & $1.294$ & $12.0$  & $1.021$ & $1.010$ \\
$0.281$ & $1.400$ & $1.302$ & $13.2$  & $0.840$ & $0.843$ \\
\tableline
\tableline
\end{tabular}
\end{center}
\end{table*}

Being a toy model, its validity is constrained to within specific
ranges of the space of parameters, which we discuss below and which
allow us nevertheless to cover essentially all of the complete space
of parameters. The first constraint on the range of validity comes
from the mass ratio, which cannot be too large since the affine model
assumes that the NS inspirals as a test fluid and is therefore
increasingly more accurate the smaller the mass ratio. As a result, we
will consider only binaries with mass ratios
\begin{equation*}
0.10 \leq q \leq 0.33\,. 
\end{equation*}
While this condition removes several of the values reported
in~\citet{Shibata:2009cn}, it is not at all unrealistic. We recall, in
fact, that the most recent estimates for the mass accreted onto the
primary compact object during the common-envelope phase are rather low
and thus the BH masses in close BH-NS binaries are likely to fall
primarily in values near $M_\text{BH} \simeq
10\,M_\odot$~\citep{Belczynski07}. Considering a canonical
$1.4\,M_{\odot}$ NS, BH-NS systems are therefore most likely to come
in a mass ratio that is $q \simeq 0.14$.

\noindent The second constraint comes from the stellar compactness,
which cannot be too small for a relativistic compact star, nor too
large given the test-fluid hypothesis of the affine model. As a
result, we will consider only binaries where the NS has
\begin{equation*}
0.1 \leq C \leq 0.16\,.
\end{equation*}
This range covers well the one considered so far in numerical
simulations (\cf Table~\ref{tab:tune}), but it is worth remarking that
the recent arguments made in~\citet{Ozel:2010} suggest a rather high
and generic compactness, $C\sim 0.16$, which is at the edge of the
range considered here, and that a standard cold EOS, such as the
APR~\citep{Akmal1998a} EOS, leads on average to compactnesses $C \sim
0.18$, thus outside of the range considered here. Future simulations
in which this EOS is employed will help extend the range of validity
in compactness of the toy model.

\noindent The third constraint comes from the BH spin, which we cannot
take as too large given that we treat the motion of the disrupted NS
with geodesics and these would lead to incorrect results if the
dimensionless spin parameter $a\equiv J/M^2_{\text{BH}}$ is too high
(\eg the ratio $M_{b, \text{tor}}/M_{b, \text{NS}} \rightarrow 1$ for
$a\rightarrow 1$). As a result, we will consider only binaries where
the BH has
\begin{equation*}
0.0 \leq a \leq 0.85\,.
\end{equation*}

\noindent The fourth and final constraint comes from the mass of the
torus, for which we need a lower limit. This is even true for
numerical simulations, whose precision is not infinite. As a
result, we consider the tori to have a zero baryon mass if
\begin{equation*}
M_{b,\text{tor}} \leq 0.01\,M_{b,\text{NS}} \simeq 0.014\,M_{\odot}\,.
\end{equation*}

The results of the comparison are summarized in Table \ref{tab:tune},
where, in addition to our numerical simulations~\citep{Tonita:2010},
we have considered also the data reported in~\citet{Duez09,
  Etienne:2008re}, and~\citet{Shibata:2009cn}. The parameters of each
BH-NS binary are reported in the first four columns of the table:
these are the adiabatic index $\Gamma$ of the NS EOS, the NS
compactness $C$, the binary mass ratio $q$, and the initial BH spin
$a$, which does not change from its initial value in our toy
model. The last three columns provide, instead, the baryonic torus
masses $M_{b, \text{tor}}$ obtained from the toy model or by the fully
general-relativistic calculations (both in units of the NS baryonic
mass $M_{b, \text{NS}}$), and the relative percentage error.

A rapid inspection of the Table and in particular of its last column
clearly shows that there are four out of sixteen numerical relativity
results that have rather large errors, \ie between $50$ and
$100\%$. Not having a clear measure of the error associated with those
simulations, it is hard to judge whether this is a limit of the toy
model or whether this is a limit of the numerical simulations. It
should be remarked, however, that these simulations are those which
report tori masses that are close to the limit we consider reasonable
(\ie $M_{b,\text{tor}}/M_{b,\text{NS}} \simeq 0.01$) and clearly new
simulations of those binaries are necessary to settle these
differences. However, with the exception of those cases, the Table
also reveals that the toy model can reproduce the remaining cases
(which represent three quarters of the data available) with an error
which is at most $18\%$ and is virtually $0$ for most of the
cases. Considering that the numerical-relativity simulations in
Table~\ref{tab:tune} were performed with different codes, different
initial separations and amounts of eccentricity\footnote{Although not
  often discussed, the presence of eccentricity in the initial data
  can lead to significant changes in the mass of the torus in BH-NS
  mergers~\citep{Tonita:2010}.}, we believe that the tuning made for
the toy model is both reasonably robust and accurate.

\begin{figure}[t]
  \begin{center}
    \includegraphics[width=8.0cm,angle=-0]{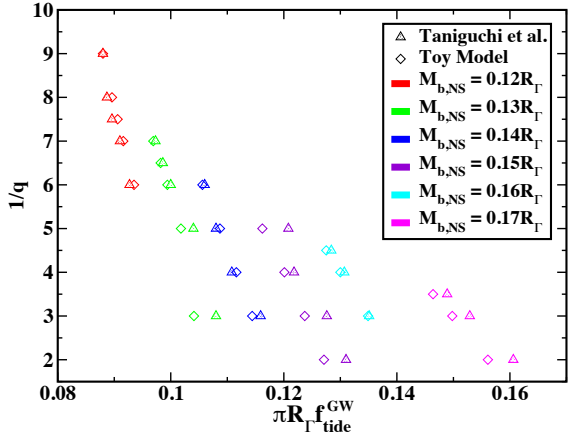}
    \caption{Gravitational-wave frequency at tidal disruption shown as
      a function of the (inverse of the) mass ratio $q$ (here
      $R_\Gamma$ is the polytropic length scale
      $K^{1/(2\Gamma-2)}$). Data plotted with diamonds is produced
      with our toy model, while data plotted with triangles is taken
      from \citet{Taniguchi:2008a}. All neutron stars have $\Gamma=2$,
      while their baryonic mass is indicated by the colour code in the
      legend. This figure should be compared with the corresponding
      Fig.~1 in \citet{Ferrari09}. \label{fig:TBFS_Cfr}}
  \end{center}
\end{figure}

To further validate the model, we use it to determine the frequency of
the gravitational radiation emitted at the onset of the tidal
disruption. Results for these frequencies were provided, for example,
by \citet{ShibataTaniguchi2008} and~\citet{Taniguchi:2008a}, and we
compare to the former in Table~\ref{tab:freq} and to the latter in
Fig.~\ref{fig:TBFS_Cfr}. The frequencies given in
\citet{Taniguchi:2008a} are found by using quasi-equilibrium sequences
of mixed binaries in circular orbits, obtained by solving the Einstein
constraint equations in the conformal thin-sandwich decomposition and
the relativistic equations of hydrostationary equilibrium; a fitting
formula for the frequency is also provided and this is used, in turn,
in~\citet{ShibataTaniguchi2008}. All cases considered by these authors
refer to nonspinning BHs and irrotational NSs, so that, for our model,
the gravitational-wave frequency at tidal disruption is given by [\cf
eq.~(\ref{eq:dthetadt})]
\begin{eqnarray}
f_\text{tide}^\text{GW}=\frac{1}{\pi}\sqrt{\frac{M_\text{BH}}{r_\text{tide}^3}}\,,
\end{eqnarray}
\ie by twice the Schwarzschild orbital frequency at the tidal
disruption radius. We note that Fig.~\ref{fig:TBFS_Cfr} reports the
(inverse of the) mass ratio $q$ as a function of $\pi R_\Gamma
f_\text{tide}^\text{GW}$, where $R_\Gamma$ is the polytropic length
scale $K^{1/(2\Gamma-2)}$. Data plotted with diamonds is produced with
our toy model, while data plotted with triangles is taken
from~\citet{Taniguchi:2008a}. All neutron stars have $\Gamma=2$, while
their baryonic mass is indicated by the colour code in the legend. An
inspection of Table~\ref{tab:freq} and of Fig.~\ref{fig:TBFS_Cfr}
shows that a very good agreement is obtained not only for the torus
mass, but also for the gravitational-wave frequency. Stated
differently, the assumptions that go into our toy model allow us to
accurately capture both the orbital evolution soon \textit{before} the
NS disruption takes place and the dynamics of the matter
\textit{after} the NS has been disrupted. Moreover, a comparison
between our Fig.~\ref{fig:TBFS_Cfr} and Fig.~1 in~\citet{Ferrari09}
shows that the present implementation of the affine model is
significantly improved with respect to its quasi-equilibrium
formulation.

Before concluding this Section, it is important to note that the
choice of a critical value for the axis ratio
$(a_2/a_1)_{\text{crit}}$ also allows us to determine the ratio
between the NS self-gravity and the tidal forces. Using a well-known
Newtonian argument, when the binary is at the separation
$r_{\text{tide}}$, the ratio between the tidal and the NS
self-gravitational force for a fluid element on the stellar surface
when the binary is at the separation $r_{\text{tide}}$ is
\begin{equation}
  \frac{M_{\text{BH}}}{M_{\text{NS}}}
  \left(\frac{a_1}{r_{\text{tide}}}\right)^3 = {\cal R}\,,
\end{equation}
so that when ${\cal R}=1$ the tidal and gravitational forces are
equal. Using the affine model and considering the tidal radius as the
one at which $a_2/a_1=(a_2/a_1)_{\text{crit}}=0.44$, we can compute
the values of the two forces at tidal disruption. Doing so for the
binaries considered in Table~\ref{tab:tune}, we find that ${\cal
  R}\simeq 0.59-0.70$ for the $\Gamma=2$ cases and ${\cal R}\simeq
0.46-0.47$ for the $\Gamma=2.75$ cases. Our tuning thus reveals that
the tidal disruption begins earlier than one would naively think and
when the tidal force is only $\sim 1/2-1/3$ the self-gravitational
one. The tidal force when the binary is at the separation
$r_{\text{tide}}$ may also be compared to the self-gravitational force
of the star at isolation, \ie when it is a sphere of radius
$R_\text{NS}$. This amounts to calculating the ratio $\cal{R'}$
when $a_1 \rightarrow R_{\text{NS}}$, \ie
\begin{equation}
\frac{M_{\text{BH}}}{M_{\text{NS}}}
\left(\frac{R_\text{NS}}{r_{\text{tide}}}\right)^3 = {\cal R'}\,,
\end{equation}
and enables us to compare our results with those
of~\citet{Taniguchi:2008a}, where it was found that ${\cal{R'}}\simeq
0.07$. More specifically, for the binaries considered in
Table~\ref{tab:tune} we find ${\cal{R'}}\simeq 0.08-0.11$, which is in
good agreement with the aforementioned result. Stated differently,
this reveals that the tidal disruption begins when the tidal force is
roughly only $\sim 1/10$ of the NS self-gravity at infinite
separation.

\section{Results}\label{sec:results}

Having tuned and validated the model, we will next consider its
predictions for the baryon mass of the torus as a function of the mass
ratio $q$, the stellar compactness $C$, and the BH spin $a$. Because
this space of parameters is three-dimensional, it is more convenient to
consider constant-spin slices and hence we will first comment on a
fiducial case of a spinning BH with $a=0.4$ and then discuss how these
results change across the possible values of the spin. 

\begin{figure}[b]
  \begin{center}
    \includegraphics[width=8.0cm,angle=-0]{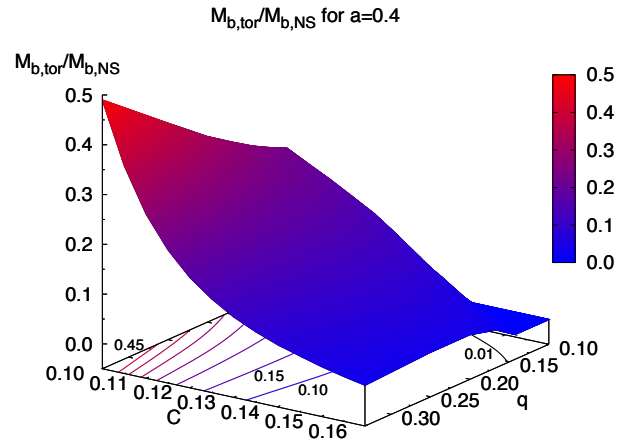}
    \vskip 1cm
    \caption{\label{fig:a04-3D}Baryonic torus mass in units of the
      stellar mass $M_{b,\text{tor}}/M_{b,\text{NS}}$ shown as a
      function of the compactness $C$ and of the mass ratio $q$, for a
      BH with spin parameter $a=0.4$.}
  \end{center}
\end{figure}

Most of our results are summarized in Fig.~\ref{fig:a04-3D}, which
shows the baryonic mass of the torus in units of the stellar mass,
$M_{b,\text{tor}}/M_{b,\text{NS}}$, as a function of the stellar
compactness $C$ and of the binary mass ratio $q$, with the data
referring to a binary in which the BH has a dimensionless spin parameter
$a=0.4$. Quite clearly, the final mass in the torus varies
considerably across the possible space of parameters and is
systematically larger the smaller $C$. This is rather obvious: the
smaller the compactness, the more ``Newtonian'' the star will be and
thus with a smaller effective gravity at the surface. In turn, this
means that, all else equal, it will be easier to disrupt it even at
large distances from the BH (\ie $r_{\text{tide}}$ is comparatively
large) and hence to produce a larger torus.

\begin{figure*}[ht]
  \begin{center}
    \includegraphics[width=8.0cm,angle=-0]{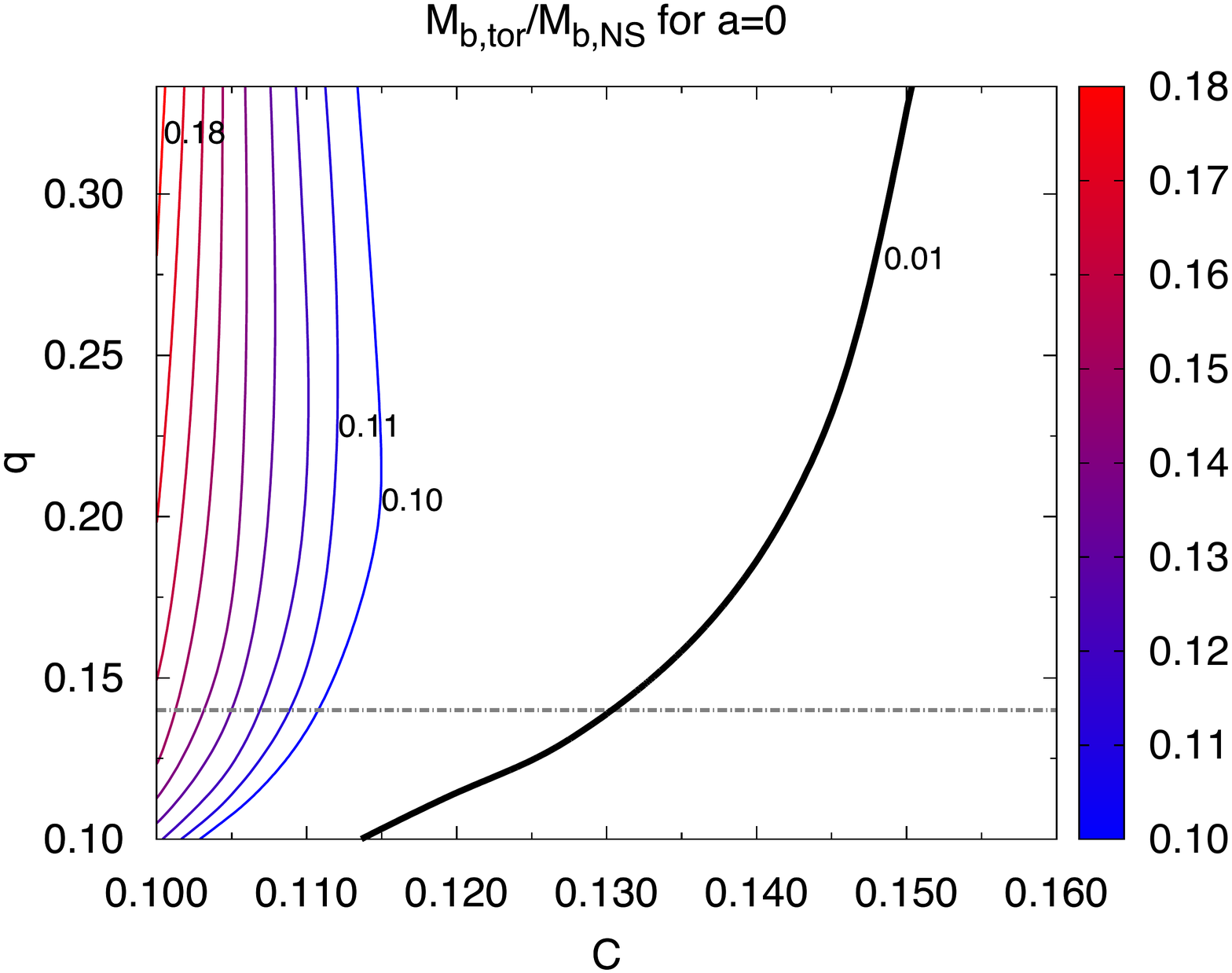}
    \hskip 1.cm
    \includegraphics[width=8.0cm,angle=-0]{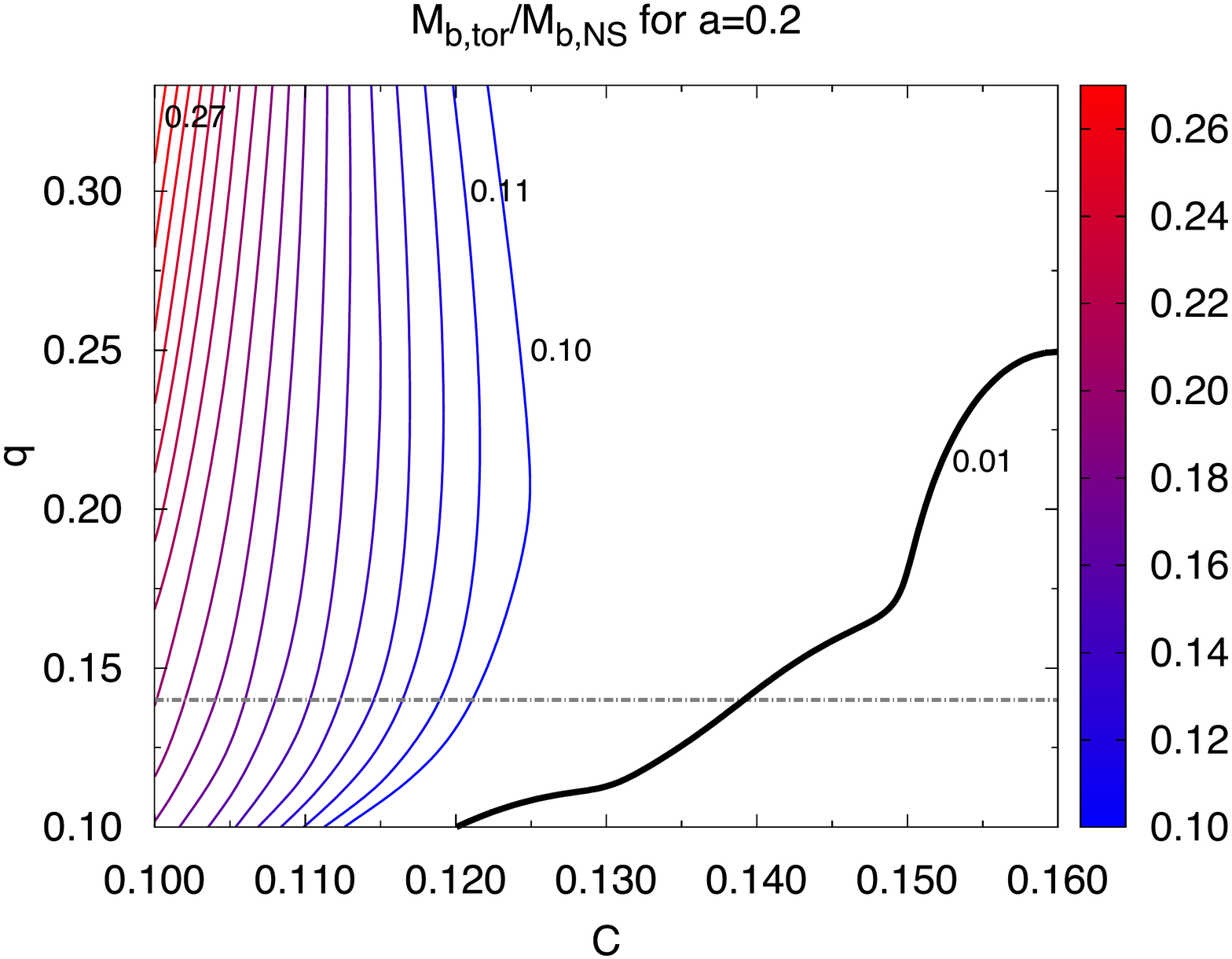}
    \vskip 1.5cm
    \includegraphics[width=8.0cm,angle=-0]{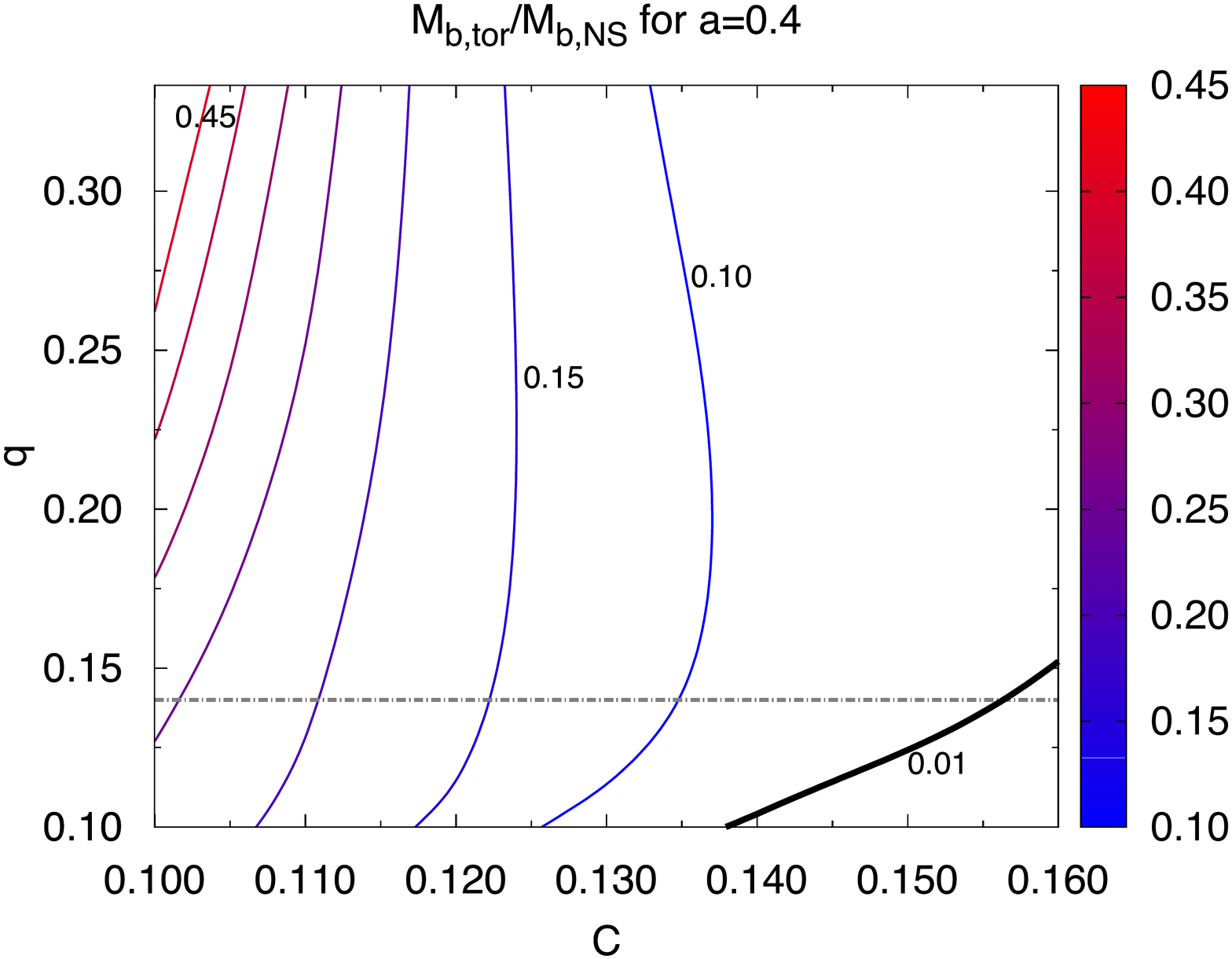}
    \hskip 1.cm
    \includegraphics[width=8.0cm,angle=-0]{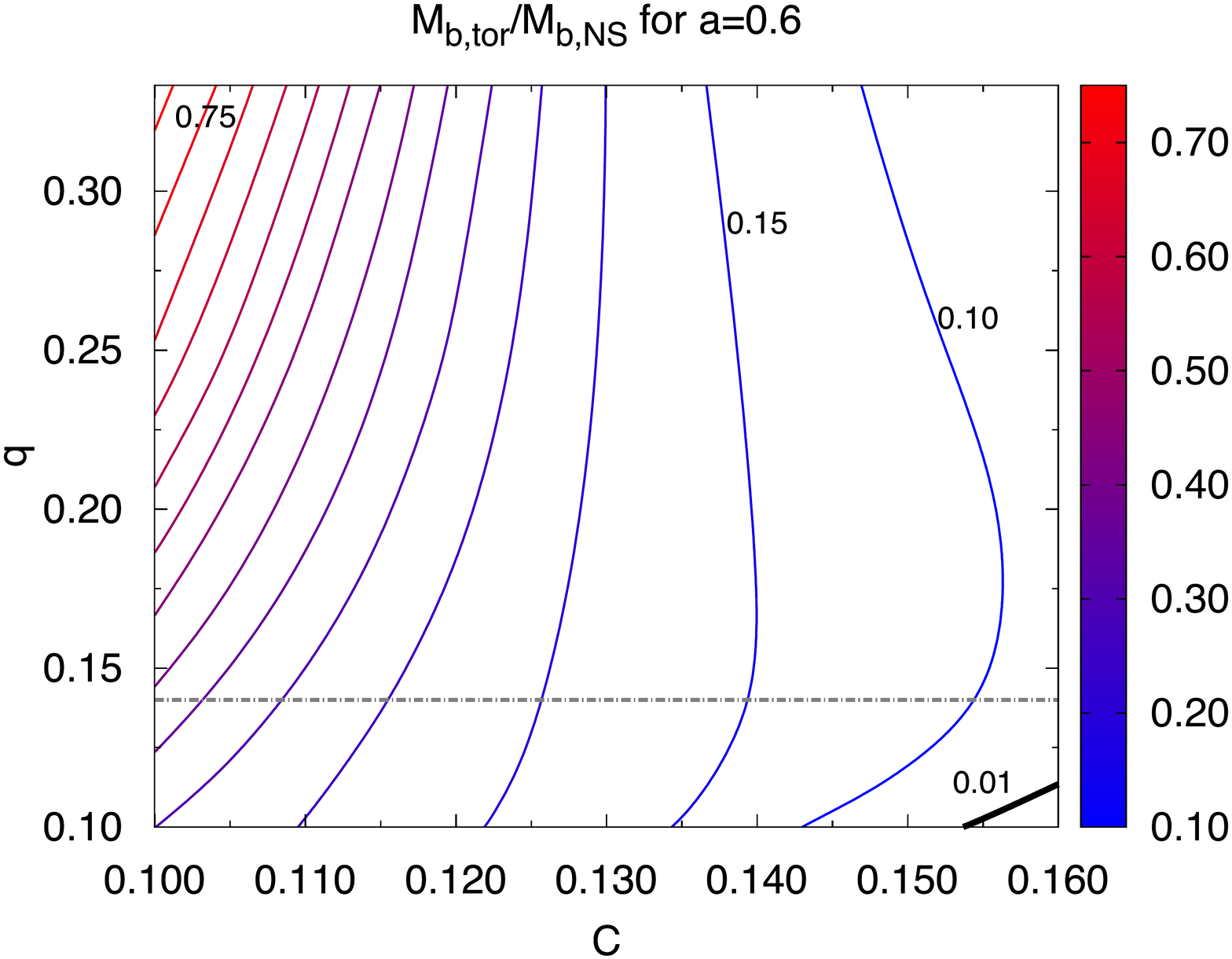}
    \vskip 1.5cm
    \includegraphics[width=8.0cm,angle=-0]{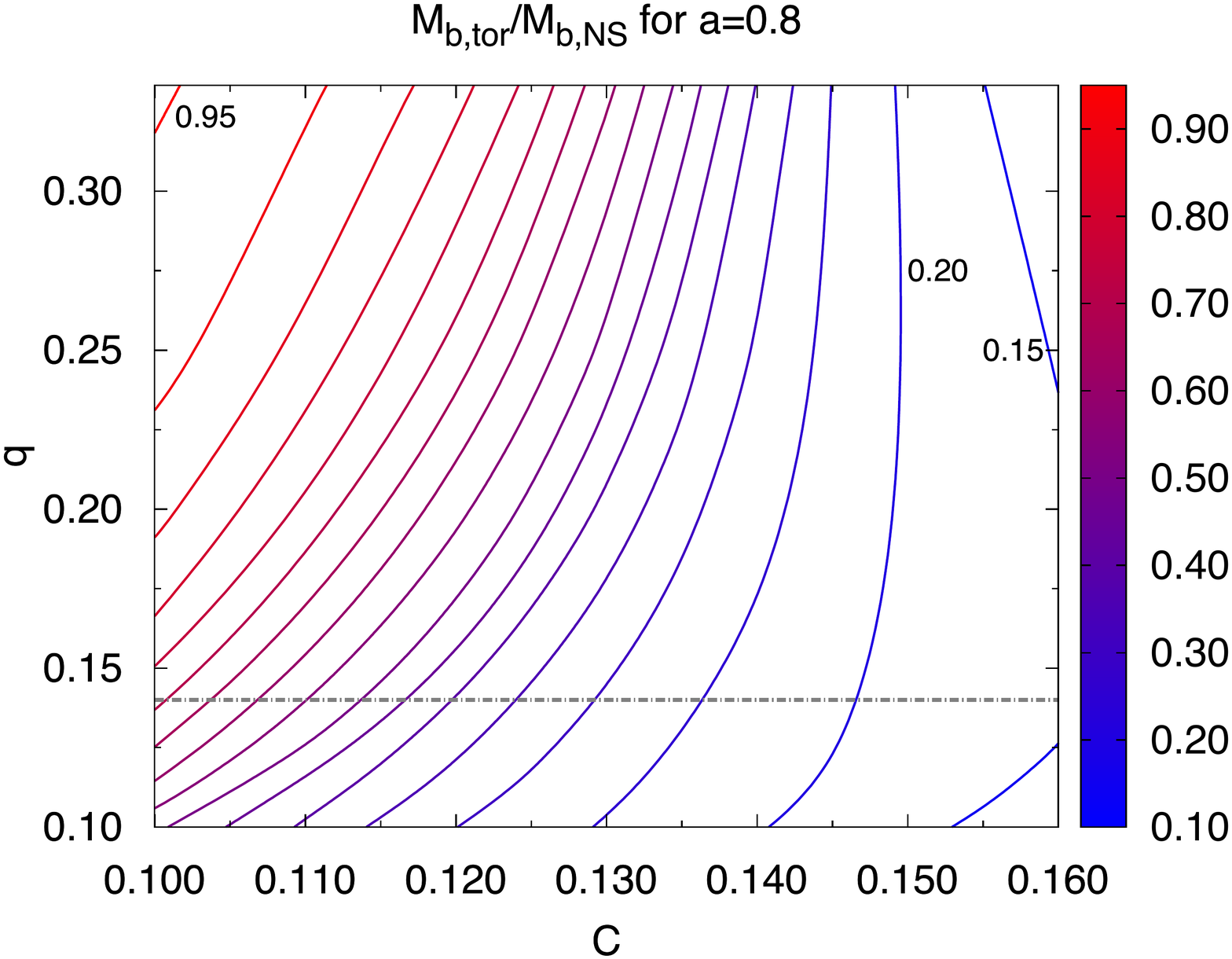}
    \hskip 1.cm
    \includegraphics[width=8.0cm,angle=-0]{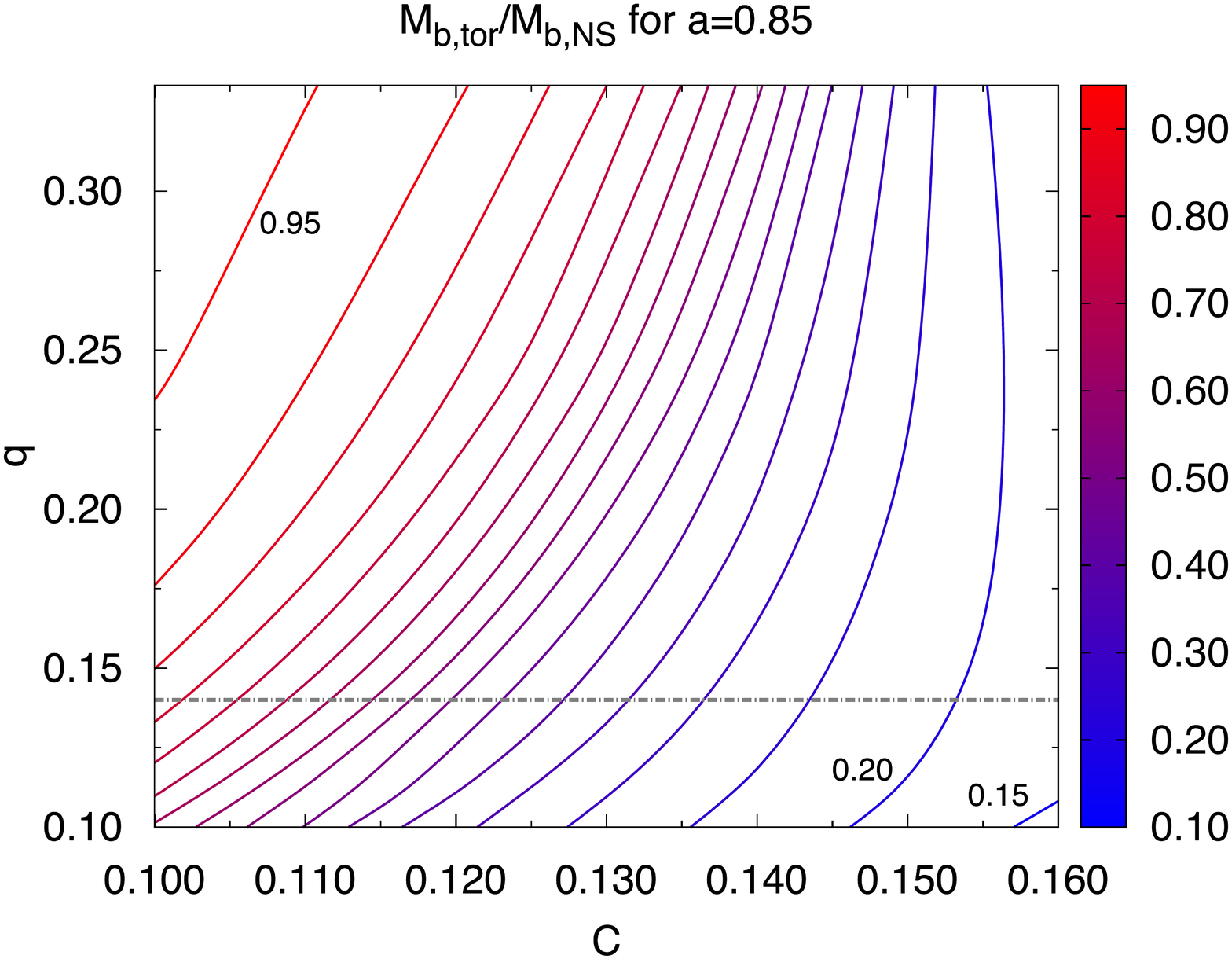}
    \vskip 1cm
    \caption{\label{fig:MT_vs_a}Baryonic torus mass in units of the
      stellar mass $M_{b,\text{tor}}/M_{b,\text{NS}}$ shown as a
      function of the compactness $C$ and of the mass ratio $q$. From
      top to bottom the different panels refer to different values of
      the BH spin ($a=0.0, 0.2, 0.4, 0.6, 0.8, 0.85$) and the numbers
      on the iso-mass contours indicate the constant spacing. Shown
      with a thick and black solid line is the area below which no
      torus is created (\ie the ``no-torus'' region with
      \hbox{$M_{b,\text{tor}}/M_{b,\text{NS}} < 0.01$}), while shown
      with a horizontal dot-dashed line is the most-likely mass ratio
      for a canonical $1.4\,M_{\odot}$ NS.}
  \end{center}
\end{figure*}

At the same time, Fig.~\ref{fig:a04-3D} shows that the mass in the
torus will be larger when the BH and the NS have comparable
masses. Also this result is quite obvious: the smaller the mass ratio,
the more unlikely it will be for the star to be tidally disrupted and
to be accreted ``whole'' by the BH. Putting things together, a BH-NS
system with large mass ratio and small compactness maximizes the
yields in terms of torus mass. For the same reasons, a binary with
with small mass ratio and large compactness will yield the smallest
tori. To fix the ideas: for a BH-NS system with $a=0.4$, the toy model
predicts that $M_{b,\text{tor}}/M_{b,\text{NS}} \sim 0.5$ when
$C=0.10$ and $q=0.33$, while essentially no tori are produced for
$q\lesssim 0.14$ and $C \gtrsim 0.14$ (\cf third panel of
Fig.~\ref{fig:MT_vs_a} where this data is also shown with contour
plots). Overall, our toy model suggests that at least statistically
\textit{a BH with spin larger than $\simeq 0.4$ is necessary to
  produce any astrophysically relevant torus}.

\begin{figure*}[ht]
  \begin{center}
    \includegraphics[width=8.0cm,angle=-0]{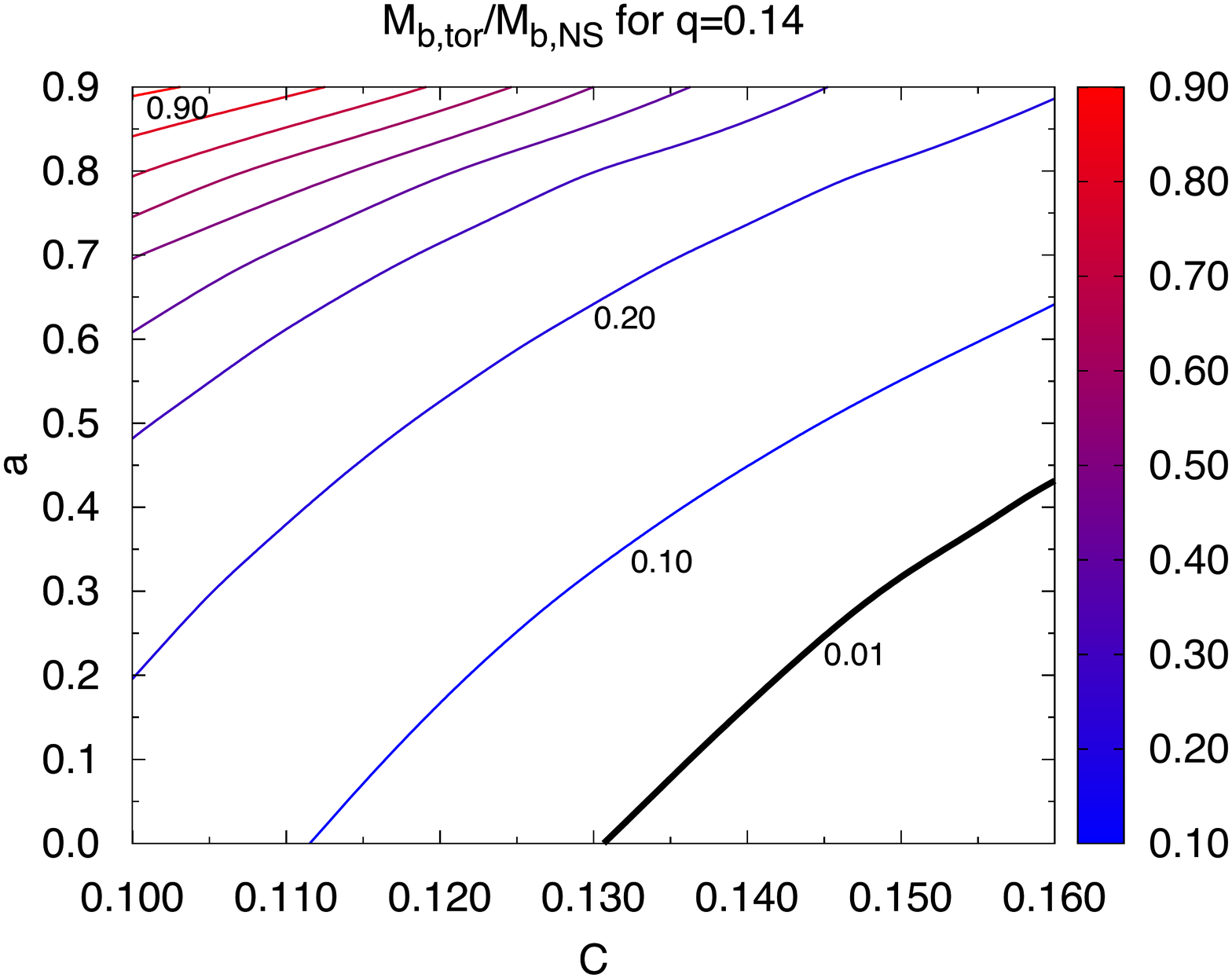}
    \hskip 1cm
    \includegraphics[width=8.0cm,angle=-0]{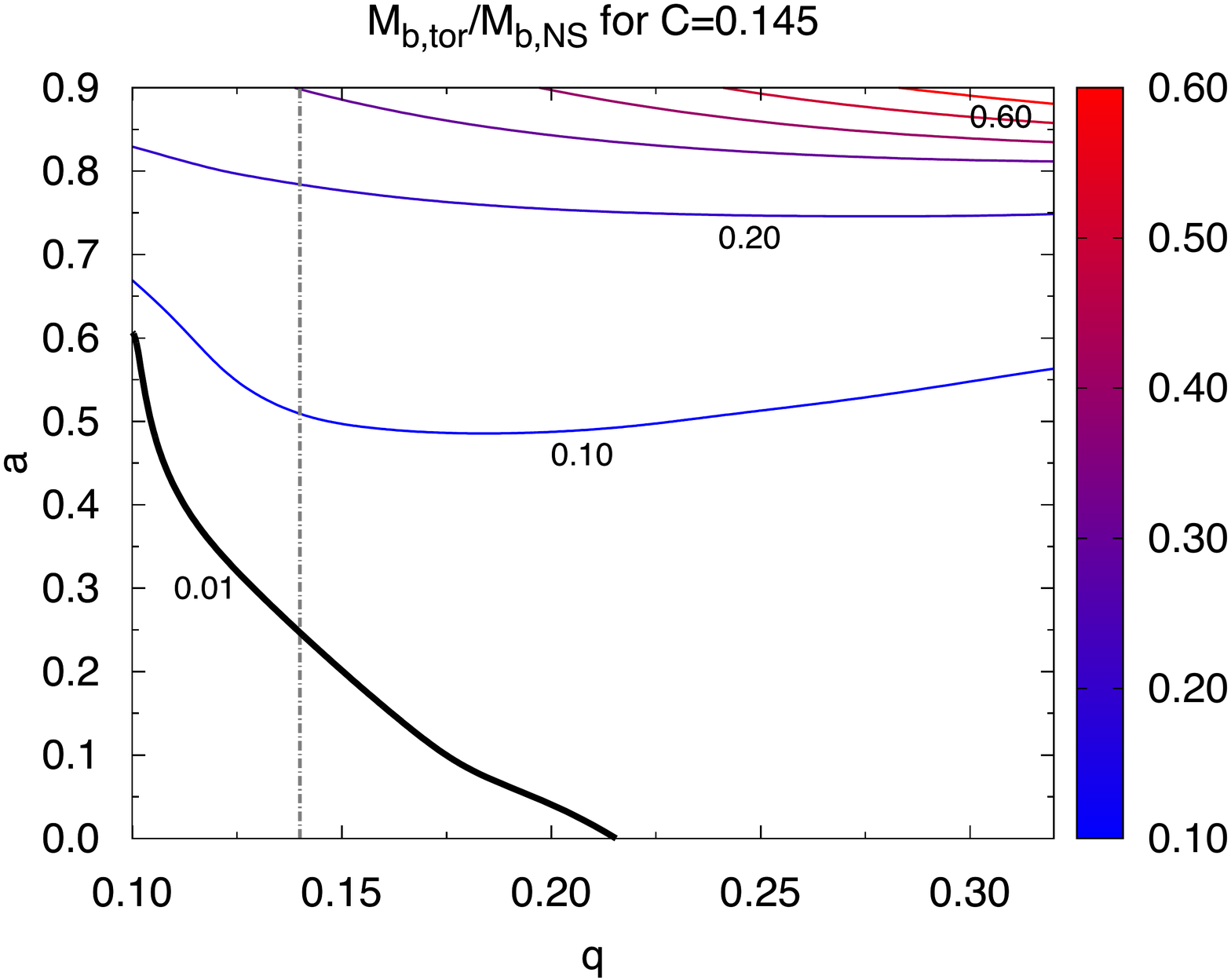}
    \vskip 1cm
    \caption{\label{fig:C0145q014} The same as in
      Fig.~\ref{fig:MT_vs_a} but considering different slices of the
      space of parameters. In particular, the left panel shows the
      baryonic torus mass in units of the stellar mass
      $M_{b,\text{tor}}/M_{b,\text{NS}}$ as a function of the BH spin
      and mass ratio for $C=0.145$, while the right panel shows
      baryonic torus mass as a function of BH spin and compactness
      for $q=0.14$.}
  \end{center}
\end{figure*}

The generic predictions of the toy model for $a=0.4$ remain unchanged
when considering also other BH spins, extending smoothly from smaller
to larger spins. This is summarized in Fig.~\ref{fig:MT_vs_a}, whose
different panels from top to bottom refer to $a=0.0, 0.2, 0.4, 0.6,
0.8, 0.85$, respectively. The baryonic mass of the torus is still in
units of the NS mass and is reported as a function of the compactness
and of the mass ratio, but it is shown through contour plots to
better quantify the results. The numerical values of some
representative contour lines are shown and allow for a direct
measurement (the contours are equally spaced in a linear scale), while
the thick and black solid line shows the area below which no torus is
created (\ie the ``no-torus'' area with
\hbox{$M_{b,\text{tor}}/M_{b,\text{NS}} < 0.01$}). Finally, shown with
a horizontal dot-dashed line is the most likely mass ratio for a
canonical $1.4\,M_{\odot}$ NS.

Moving from the top to the bottom of Fig.~\ref{fig:MT_vs_a} it is easy
to recognize that the maximum mass attained at the smallest
compactness increases significantly with the BH spin, ranging from
$M_{b,\text{tor}}\simeq 0.18\,M_{b,\text{NS}} \simeq 0.25\,M_{\odot}$
for $a=0.0$, to $M_{b,\text{tor}} \gtrsim 0.95\,M_{b,\text{NS}} \simeq
1.33\,M_{\odot}$ for $a=0.85$. At the same time, the ``no-torus'' area
decreases and virtually disappears for $a \gtrsim 0.6$. Stated
differently, for sufficiently large BH spins a torus is
\textit{always} produced and with non-negligible mass. As an example,
taking as fiducial compactness the canonical value of $C\simeq 0.145$,
the torus mass at the fiducial mass ratio goes from
$M_{b,\text{tor}}\simeq 0.06\,M_{b,\text{NS}} \simeq 0.08\,M_{\odot}$
for $a=0.4$, to $M_{b,\text{tor}} \simeq 0.24\,M_{b,\text{NS}} \simeq
0.34\,M_{\odot}$ for $a=0.85$.

A complementary view, because it refers to a different slicing of the
space of parameters, is illustrated in Fig.~\ref{fig:C0145q014}, which
is the same as in Fig.~\ref{fig:MT_vs_a}, but it shows the baryonic
torus mass as a function of the BH spin and of the mass ratio for a
fixed compactness $C=0.145$ (left panel), or as a function of the BH
spin and of the compactness for a fixed mass ratio $q=0.14$ (right
panel). Both panels of the figure are rather self-explanatory and
underline what has already been discussed above: large tori masses are
possible for BHs which are spinning sufficiently rapidly or for neutron 
stars which are not very compact (favoring stiff equations of state).

In summary: considering an astrophysically realistic mass ratio $q
\simeq 0.14$ and a conservative value of the stellar compactness
$C\simeq 0.145$ (we recall that even larger values were recently
suggested in~\citet{Ozel:2010}), the predictions of the toy model are
that the torus mass should be
\begin{equation}
M_{b,\text{tor}} \lesssim 0.24\,M_{b,\text{NS}} \simeq 0.34\,M_{\odot}\,,
\end{equation}
for BH spins $0 \leq a \leq 0.85$. Such masses are comparable but also
smaller than the ones predicted by the analysis of unequal-mass NS-NS
mergers carried out by~\citet{Rezzolla:2010}.

\section{An intuitive interpretation}\label{sec:intepretation}
In the previous Section we have shown that the complex dynamics of the
tidal disruption and subsequent accretion onto the BH is well-captured
by the simple assumptions needed to build our toy model. In what
follows we will show that an even simpler framework can be built to
explain at least qualitatively the results of the toy model.

We have already noted that binaries with less compact NSs produce
bigger tori as these are more ``Newtonian'' and hence can sustain
smaller tidal forces before being disrupted. Stated differently, less
compact stars have larger tidal radii $r_\text{tide}$. In
the usual arguments this quantity is generally compared to the
ISCO, and the standard line of arguments says that a BH-NS binary will
produce a torus if $r_\text{tide} \gtrsim r_\text{ISCO}$. This
reasoning, however, is inadequate for systems which yield low mass tori. 
An obvious failure of the argument
is offered by a NS that is disrupted by its BH companion exactly at
$r_\text{tide}=r_\text{ISCO}$. In this case half of the star would
still be outside the ISCO and thus potentially capable of producing a
torus.

The necessary, but not sufficient, condition for a BH-NS binary to
yield a torus is thus better expressed as
\begin{equation}
\label{eq:clear}
\frac{r_\text{tide}+a_1(r_\text{tide})-r_\text{ISCO}}{2R_\text{NS}} \simeq
1 + \frac{r_\text{tide}-r_\text{ISCO}}{2R_\text{NS}} > 0\,,
\end{equation}
where the second expression is obtained after recognizing that at
tidal disruption $a_1(r_\text{tide})\simeq
2R_\text{NS}$. Expression~\eqref{eq:clear} has three important
properties: it is dimensionless, it combines the three fundamental
lengthscales of our system, and it essentially measures how many NS
diameters fit between $r_{\text{tide}}+a_1(r_\text{tide})$ and
$r_{\text{ISCO}}$ (\cf Fig.~\ref{cartoon}). In other
words,~\eqref{eq:clear} quantifies ``how much useful space'' there is
for the NS to form a torus after it is tidally disrupted.

At this point it is natural to associate this quantity directly to the
mass of the torus as expressed in units of the NS mass
\begin{equation}
\label{eq:explain}
\frac{M_{b,\text{tor}}}{M_{b,\text{NS}}} \propto \left[ 1 + 
\frac{r_\text{tide}-r_\text{ISCO}}{2R_\text{NS}}\right]\,,
\end{equation}
where the exact proportionality will depend (albeit weakly) on $q$ and
$C$. Not surprisingly,~\eqref{eq:explain} reproduces, at least
qualitatively, all of the phenomenology discussed before and predicted
by our toy model. As an example, for a fixed BH spin, and hence fixed
$r_\text{ISCO}$, the torus mass will increase for less compact NSs
since for these $r_\text{tide}$ will grow and more rapidly than
$R_{\text{NS}}$. Similarly, for a fixed compactness, and hence for a
fixed $R_\text{NS}$, the torus mass will grow with the BH spin as does
the difference $r_\text{tide} - r_{\text{ISCO}}$ ($r_{\text{ISCO}}$
decreases more rapidly than $r_{\text{tide}}$).

\section{Concluding remarks}\label{sec:concl}

The production of a massive torus orbiting stably around a rotating BH
is a necessary ingredient in all models that explain SGRBs in terms of
the coalescence of binary systems composed of a BH and a NS or of two
NSs. The accurate calculation of this mass inevitably requires the use
of numerical-relativity simulations, which however are still very
expensive and have so far been applied only to a tiny patch of the
possible space of parameters. In the case of BH-NS binaries
especially, the space of parameters is particularly extended as it
involves the mass ratio of the binary $q$, the stellar compactness
$C$, and the BH spin $a$. As a result, we presently have only a very
limited idea of what are the likely torus masses that this process
will yield and hence have a rather limited ability to assess whether
or not the merger of a BH-NS system under astrophysically realistic
conditions will serve as a robust scenario for the powering of SGRBs.

To compensate for this lack of knowledge, we have developed a toy
model that allows for the computation of the mass of the torus without
having to perform a numerical-relativity simulation. In essence, we
model the NS in the binary as a tri-axial ellipsoid which is tidally
distorted as it orbits in the tidal field of a rotating BH and as
described by the relativistic affine model. When the star is
disrupted, we decompose it into a system of non-interacting ``fluid
particles'' which move on the corresponding geodesics. We therefore
compute the mass of the torus as the integral of the masses of the
particles which do not fall into the BH. The only free parameter in
our model is the radius at which the tidal disruption takes place and
which we tune in terms of the ratio of the semi-major axes on the
equatorial plane and with the aid of numerical-relativity
simulations. The tuning requires care, but allows us to reproduce with
precision the majority of the data, some of which shows
inconsistencies of their own.

As it is natural for a semi-analytic approach, the model has a limited
range of validity, which we have decided to set in the following
ranges for the mass ratio, compactness and BH spin: $0.10\leq q\leq
0.33$, $0.1\leq C\leq 0.16$, $0.0\leq a\leq 0.85$, respectively.
Overall, the toy model predicts that high BH spins, small mass ratios
and small NS compactnesses all enhance the mass of the remnant
torus. As a result, tori with masses as large as $M_{b,\text{tor}}
\simeq 1.33\,M_{\odot}$ are predicted for a very extended star with a
stellar compactness $C\simeq 0.1$ inspiralling around a BH with
dimensionless spin $a=0.85$ and mass ratio $q\simeq 0.3$.
However, when considering a more astrophysically reasonable mass ratio
$q \simeq 0.14$ and a conservative but realistic value of the
compactness $C\simeq 0.145$, the predictions of the toy model set a
considerably smaller upper limit of $M_{b,\text{tor}} \lesssim
0.34\,M_{\odot}$.

All of the phenomenology discussed above has a rather intuitive
interpretation and it is easy to show that the torus mass is directly
related to how much of the star falls between the tidal radius
augmented of the NS semi-major axis and the ISCO. Hence, collecting
the three fundamental lengthscales appearing in the system, the simple
expression ${M_{b,\text{tor}}}/{M_{b,\text{NS}}} \propto [ 1 +
  ({r_\text{tide}-r_\text{ISCO}})/2R_\text{NS}]$ is able to capture
qualitatively the predictions that our toy model can make
quantitatively.

Toy models are by construction approximate representations of much
more complex phenomena and their predictions are therefore
intrinsically accompanied by errors. Bearing this in mind, the toy
model presented here can be further improved as new and more accurate
numerical-relativity simulations are performed and as their level of
realism increases with the inclusion of microphysical EOSs, magnetic
fields and radiative transfer. This will represent the focus of our
future work.

\acknowledgments It is a pleasure to thank Bruno Giacomazzo and
Jocelyn Read for useful discussions, and William Lee and Max Ruffert
for their comments. This work was supported in part by the IMPRS on
``Gravitational-Wave Astronomy'', by the DFG grant SFB/Transregio~7,
and by ``CompStar'', a Research Networking Programme of the European
Science Foundation.


\appendix
\section{The Parallel Propagated Tetrad}
\label{app:parallel-tetrad}

In this Appendix, we gather together the equations derived
by~\citet{Marck83} to define a tetrad which is parallel-transported as
it moves along a timelike geodesic of a Kerr BH spacetime in
Boyer-Lindquist coordinates
\begin{align}
ds^2&=-\left(1-\frac{2M_{\text{BH}}r}{\Sigma}\right)\textrm{d}t^2
-\frac{4M_{\text{BH}}r}{\Sigma}{\tilde a}\sin^2\theta\textrm{d}t\textrm{d}\phi
+\frac{\Sigma}{\Delta}\textrm{d}r^2 + \Sigma\textrm{d}\theta^2
+\frac{\mathcal{A}}{\Sigma}\sin^2\theta d\phi^2\,,
\end{align}
where
\begin{align}
\Sigma &\equiv r^2 + {\tilde a}^2\cos^2\theta\,,\\
\Delta &\equiv r^2 + {\tilde a}^2 - 2M_{\text{BH}}r\,,\\
\mathcal{A} &\equiv (r^2+{\tilde a}^2) - \Delta {\tilde a}^2\sin^2\theta\,.
\end{align}
In our model, the test particle is identified with the centre of mass
of a NS orbiting its rotating BH companion. Marck expresses his
results in the canonical symmetric orthonormal tetrad introduced
by~\citet{Carter68}
\begin{align}
\boldsymbol{\omega}^{(0)} &= \sqrt{\frac{\Delta}{\Sigma}} (\textrm{d}t -
{\tilde a}\sin^2\theta \textrm{d}\phi)\,,\\
\boldsymbol{\omega}^{(1)} &= \sqrt{\frac{\Delta}{\Sigma}} \textrm{d}r\,,\\
\boldsymbol{\omega}^{(2)} &= \sqrt{\Sigma} \textrm{d}\theta\,,\\
\boldsymbol{\omega}^{(3)} &= \frac{\sin\theta}{\sqrt{\Sigma}} [{\tilde a}\textrm{d}t -
(r^2+{\tilde a}^2)\textrm{d}\phi]\,,
\end{align}
which has the convenience of casting the Kerr metric in the form
\begin{align}
ds^2 = \eta_{(\mu)(\nu)}\boldsymbol{\omega}^{(\mu)}
\boldsymbol{\omega}^{(\nu)}\,,
\end{align}
where $\eta_{(\mu)(\nu)}=\textrm{diag}(-1,1,1,1)$ is the metric tensor
of Minkowksi spacetime. Before expressing the components of the basis
vectors of the tetrad found in~\citet{Marck83}, we define the
quantities $\alpha$ and $\beta$
\begin{align}
\alpha &\equiv \sqrt{\frac{K-{\tilde a}^2\cos^2\theta}{r^2+K}}\,,\\
\beta &\equiv \sqrt{\frac{r^2+K}{K-{\tilde a}^2\cos^2\theta}}\,,
\end{align}
where $K$ is Carter's constant, and the two vectors
\begin{align}
\label{tilde-e1}
\tilde{e}_{1}^{\;\;(0)} &= \alpha\sqrt{\frac{\Sigma}{K\Delta}}r\dot r\,,\\
\tilde{e}_{1}^{\;\;(1)} &= \frac{\alpha r[E(r^2+{\tilde a}^2)-{\tilde a}L_z]}{\sqrt{K
    \Sigma\Delta}}\,,\\
\tilde{e}_{1}^{\;\;(2)} &= \frac{\beta {\tilde a}\cos\theta({\tilde a}E\sin\theta-L_z\sin^{-1}
  \theta]}{\sqrt{K\Sigma}}\,,\\
\tilde{e}_{1}^{\;\;(3)} &= \beta\sqrt{\frac{\Sigma}{K}}{\tilde a}\cos\theta\dot\theta\nonumber\,,
\end{align}
and
\begin{align}
\label{tilde-e2}
\tilde{e}_{2}^{\;\;(0)}&=\frac{\alpha r[E(r^2+{\tilde a}^2)-{\tilde a}L_z]}{\sqrt{\Sigma
    \Delta}}\,,\\
\tilde{e}_{2}^{\;\;(1)} &= \alpha\sqrt{\frac{\Sigma}{\Delta}}\dot r\,,\\
\tilde{e}_{2}^{\;\;(2)} &= \beta\sqrt{\Sigma}\dot\theta\,,\\
\tilde{e}_{2}^{\;\;(3)} &= \beta\frac{{\tilde a}E\sin\theta-L_z\sin^{-1}\theta}{
  \sqrt{\Sigma}}\nonumber\,,
\end{align}
where the dots indicate derivatives with respect to the proper time
$\tau$, and where $E$ and $L_z$ are, respectively, the energy and the
angular momentum about the axis of symmetry of the BH per unit mass of
the star. We are now ready to express -- in Carter's symmetric tetrad
-- the components of the vectors forming an orthonormal tetrad
parallel-transported along an arbitrary timelike geodesic in a Kerr
spacetime. These are
\begin{align}
\label{e0}
e_{0}^{\;\;(0)} &= \frac{E(r^2+{\tilde a}^2)-{\tilde a}L_z}{\Delta\sqrt{\Sigma}}\,,\\
e_{0}^{\;\;(1)} &= \sqrt{\frac{\Delta}{\Sigma}}\dot r\,,\\
e_{0}^{\;\;(2)} &= \sqrt{\Sigma}\dot{\theta}\,,\\
e_{0}^{\;\;(3)} &= \frac{{\tilde a}E\sin\theta-L_z\sin^{-1}\theta}{\sqrt{\Sigma}}\,,
\end{align}
\begin{align}
\label{e1}
\mathbf{e}_{1} &= \cos\Psi\tilde{\mathbf{e}}_{1} - \sin\Psi
\tilde{\mathbf{e}}_{2}\,,\\
\label{e2}
\mathbf{e}_{2} &= \sin\Psi\tilde{\mathbf{e}}_{1} + \sin\Psi
\tilde{\mathbf{e}}_{2}\,,
\end{align}
and
\begin{align}
\label{e3}
e_{3}^{\;\;(0)} &= \sqrt{\frac{\Sigma}{K\Delta}}{\tilde a}\cos\theta\dot r\\
e_{3}^{\;\;(1)} &= \frac{{\tilde a}\cos\theta[E(r^2+{\tilde a}^2)-{\tilde a}L_z]}{\sqrt{K
    \Sigma\Delta}}\,,\\
e_{3}^{\;\;(2)} &=-\frac{r({\tilde a}E\sin\theta-L_z\sin^{-1}\theta)}{\sqrt{K\Sigma}}\,,
\\
e_{3}^{\;\;(3)} &= \sqrt{\frac{\Sigma}{K}}r\dot\theta \,.
\end{align}
The rotation by an angle $\Psi$ in (\ref{e1}) and (\ref{e2}) ensures
that $\mathbf{e}_{(1)}$ and $\mathbf{e}_{(2)}$ are indeed
parallel-transported along any Kerr timelike geodesic. Finally, the
evolution of the angle $\Psi$ is governed by (\ref{eq:Psi}) for
circular equatorial orbits.

\bibliography{aeireferences}

\vfill

\end{document}